\documentclass[useAMS,usenatbib]{mn2e}

\usepackage{graphicx}

\title[Disks of S0 galaxies]{Ages and abundances in large-scale
stellar disks of nearby S0 galaxies
\thanks{Based on the observations with the Russian 6m telescope}
}
\author[O. K. Sil'chenko, I. S. Proshina, A. P. Shulga, and S. E. Koposov]
{O. K. Sil'chenko$^{1,2}$\thanks{E-mail:
olga@sai.msu.su (OKS)}, I. S. Proshina$^1$, A. P. Shulga$^1$, and
S. E. Koposov$^{3,1}$\\
$^{1}$Sternberg Astronomical Institute of the Lomonosov Moscow State University, Moscow, Russia\\
$^{2}$Isaac Newton Institute of Chile, Moscow Branch\\
$^{3}$Institute of Astronomy, Cambridge, UK}

\begin{document}

\date{Accepted ;Received}


\maketitle

\label{firstpage}

\begin{abstract}
By undertaking deep long-slit spectroscopy with the focal reducer SCORPIO of
the Russian 6m telescope, we studied stellar population properties
and their variation with radius in 15 nearby S0 galaxies sampling a wide
range of luminosities and environments.
For the large-scale stellar disks of  S0s, we have measured  SSP-equivalent
metallicities ranging from the solar one down to [Z/H]$=-0.4 - -0.7$, rather high
magnesium-to-iron ratios, [Mg/Fe]$\ge +0.2$, and mostly old SSP-equivalent ages.
Nine of 15 (60\%$\pm$13\%) galaxies have large-scale stellar disks older
than 10~Gyr, and among those we find all the galaxies which reside in
denser environments.
The isolated galaxies may have intermediate-age  stellar disks which are 7-9~Gyr old.
Only two galaxies of our sample, NGC~4111 and NGC~7332, reveal
SSP-equivalent ages of their disks of 2--3~Gyrs. Just these two young disks appear
to be thin, while the other, older disks have scale heights typical for
thick stellar disks. The stellar populations in the bulges at radii of $0.5r_e$ are on
the contrary more metal-rich than the solar $Z_{\odot}$, 
with the ages homogeneously distributed between
2 and 15~Gyr, being almost always younger than the disks. We conclude that
S0 galaxies could not form in groups at $z\approx 0.4$ as is thought now; a
new scenario of the general evolution of  disk galaxies is proposed instead.

\end{abstract}


\section{Introduction}

In the `tuning fork' galaxy classification scheme by \citet{hubble36} 
lenticular galaxies occupied an intermediate position between ellipticals
and spirals. They looked homogeneously red and smooth, as ellipticals,
implying that they contained an old stellar population, and they had at least
two large-scale structural components, centrally concentrated
spheroids (bulges) and extended stellar disks, as spirals. Hubble suggested
that the morphological-type sequence might be an evolutionary sequence
from the simplest shapes to more complex ones because his classification
was `almost identical with the path of development derived by Jeans from
purely theoretical investigations' \citep{hubble26}. Later, the opposite
evolution was supposed by the dominant cosmological paradigm: pure disks --
late-type spirals? -- formed first, and later they merged into spheroids --
bulges of early-type disk galaxies or elliptical galaxies. But in all
schemes lenticulars remain secondary products of galaxy evolution. The most
common opinion is presently that lenticular galaxies are former spirals which
lost their gas and could not produce more young stars.

The transformation of spirals into lenticulars is thought to be related to dense
environments. Indeed, within dense environments lenticulars are the dominant
galaxy population \citep{dressler,pg84}; whereas in the field the fraction of S0s
is only about 15\%, with spirals being the majority \ \citep{apm}. Within the typical group
environments the fractions of S0s and Spirals are comparable, both about
40\% -45\%\ \citep{pg84}, and in clusters the fraction of S0s may reach
60\%\ \citep{dressler}. The studies of galaxy morphologies
with the Hubble Space Telescope have shown, that the morphological-type balance
within dense environments changes abruptly at the redshifts of 0.4--0.5 \citep{morphs}.
The fraction of elliptical galaxies in clusters stays constant at $\sim$20\%-30\%\
between $z\approx 0.8$ and $z\approx 0.0$, while spiral galaxies, constituting about
50\% -70\%\ of all galaxies in high-redshift clusters, are replaced by lenticulars
at the redshift of 0.4 \citep{fasano}. According to these results, the exact place
of S0 formation at $z=0.4$ was supposed to be clusters and the proposed
mechanisms for this transformation were associated to dense environments via tidal effects
and interaction with hot intracluster medium
\citep{spitzerbaade,icke,moore96,byrdvalt,quilis,larsons0,shayatully,gunngott,cowieson}.
However, recently some additional observational
information was presented which gave evidence for equal presence of lenticulars and
spirals in the clusters and in groups at redshifts $z=0.4-0.5$.
\citet{wilman} have found that at $z=0.4$ the fraction
of S0s in groups is exactly the same as the fraction of S0s in clusters, and exceeds
significantly the fraction of S0s in the field, so they have concluded that the group
and subgroup environments were the main sites of the formation of S0 galaxies, and
clusters at $z=0.4$ had then to accrete the groups together with their S0s.
Moreover \citet{just2010} have measured strong evolution of the S0 fraction
in massive groups with galaxy velocity dispersions of $500-750$ km/s between
$z=0.4$ and $z=0.0$. So the probable place of the proposed galaxy transformation
has shifted now from cluster to groups.

The information on galaxy morphologies at even higher redshifts, $z>1-1.5$,
implies another possible way of S0 formation. Deep fields of the Hubble Space Telescope
have provided high-resolution images of high-redshift galaxies; the statistics of the
morphological types at $z>1-1.5$ demonstrate the absence of shapes typical for lower
redshifts (those of the `tuning fork') and the dominance of clumpy irregular types:
chains, nests, `head-tails', etc. \citep{vdb96,elm05,elm07}. Kinematical studies of
gas motions within these chains and nests have shown that these are massive,
of $10^{10}-10^{11}$ solar masses, gravitationally-bound
disks where star formation proceeds in clumps with typical sizes of about 1 kpc
and where the thickness of the disks matches just the clump sizes being 1--1.5 kpc \citep{genzel,natasha}.
Theoretical considerations confirm that gas-rich disks
at $z>2$ are gravitationally unstable and should fragment into clumps with typical masses of
$10^9$ solar mass \citep{noguchi,ceverino}. The observed timescales of star formation
in these clumpy disks is much less than $10^9$ years -- on average, $2\cdot 10^8$ years,
according to \citet{genzel}. It means that at $z>2$ we expect fast formation of
{\bf thick} stellar disks via effective consumption of gas by star formation during
a few $10^8$ years with subsequent feedback (stellar winds of massive stars,
supernova explosions) which should clear the disks of remaining gas and
would produce S0 galaxies which will be 10-12~Gyr old at $z=0$.

If the bulk of S0 galaxies were spirals at $z>0.4$, or only 4~Gyr ago according to the 
modern Universe LCDM timescale, then star formation proceeded in their disks only 4--5~Gyr
ago. In that case their disks should now contain stars as young as 4--5~Gyr and with the solar
magnesium-to-iron ratio. 
Stellar populations in cluster elliptical galaxies are much older than 4~Gyr 
\citep{thom2005}, and however the mean integrated colours of the nearby ellipticals
and lenticulars are the same \citep{butawil}.
\citet{larsons0} explained this controversy with the aid of age-metallicity
degeneracy: if the younger stellar population of S0s are also more metal-rich than
the stellar populations of ellipticals, then the colours might be the same.
The age-metallicity degeneracy problem can be solved to determine both metallicity and mean age
of the stellar populations in S0 disks by using e.g. Lick indices including H$\beta$
\citep{worth94} or by combining optical colours with the near infrared (NIR) data \citep{bothgregg}.
There were numerous attempts of doing so, but with controversial results.
By combining the optical colours with the NIR data \citet{bothgregg}
found that the S0 disks are younger by 3--5~Gyr than their centers (which they
called `bulges'), while \citet{pelbal} did not detect any age difference
between the bulges and the disks in their sample of S0s, and \citet{mcarthur},
among a few S0 galaxies, found 2 or 3 where the centers were prominently younger
than the outer disks. The dominant positive age gradients along the radius up to
several effective radii are found for a large sample
of field S0 galaxies by \citet{prchamb} from the grJH surface photometry.
The similar photometric study of 53 S0 galaxies in the Virgo cluster
by \citet{roediger} has revealed zero mean age gradient
along the radius and the very old centers of S0s ($\langle T \rangle =10.2 \pm 0.7$~Gyr).
The same scatter of conclusions is seen among the studies done by using
spectral line indices. The early paper by \citet{caldwell} where the centers of S0s
were assumed to be old {\it by definition}, demonstrated bluer $U-V$-colours of the
disks with respect to the Mg--$(U-V)$ metallicity sequence for old stellar populations,
so the disks were concluded to be younger than the `bulges'. Later, with the advent of Lick indices
and modern stellar population models, \citet{fish96} found the disks to be
older than the centers in a few nearby S0s. The same conclusion was reached
by \citet{bedregal}  and by \citet{spolaor} for small
samples of S0 Fornax members, including dwarfs and giants, while \citet{mehlert2003},
analysing data for Coma cluster member galaxies, did not see any age difference between
the centers and the outer parts of the galaxies in their sample. Deep spectral
observations of individual nearby S0 galaxies were also published: in NGC~3115 the disk 
is younger than the bulge (6~Gyr {\bf vs} 12~Gyr) \citep{n3115}, 
and in NGC~3384 the age trend is opposite \citep{sanchez}. However, we would note that
most of these studies probed the disk stellar populations at the radii of maximum 1--2
exponential scale lengths; the photometric study by \citet{prchamb} and the study of NGC~3115
are the only ones reaching outer disks of their S0s.

In the present work, we have measured variations of Lick indices along the radius up 
to 2--4 exponential scale lengths in 15 nearby lenticular galaxies in different environments.
By using photometric decomposition, we have extracted the regions where the bulge or disk
dominate, and have determined the mean (light-weighted or SSP-equivalent) ages, metallicities,
and magnesium-to-iron abundance ratios for the bulges and for the stellar disks.
These data are crucial to evaluate currently available scenarios of formation of S0 galaxies.
The paper is organized as follows. Section~2 presents our sample. Section~3 describes the
observations and data reduction. Section~4 reviews the details of the photometric structures and
gives our own photometric analysis of some moderately inclined  galaxies of our sample.
The central Section~5 presents the age and abundance measurements in the bulges and in the
large-scale disks of the galaxies under consideration. Sections 6 and 7 contain the discussion
and the conclusions respectively.

\section{The sample}

The sample consists of nearby lenticular galaxies for which
deep long-slit spectra have been obtained at the Russian 6m telescope
during the last five years as a part of several observational programs. The main body
of the sample is the set of edge-on lenticular galaxies selected for the kinematical study
and dynamical modelling by Natalia Sotnikova and observed in the frame of her
observational proposal; the kinematical profiles for these galaxies are to
be published later. For the purpose of the present work we have taken 
the raw spectral data and have derived Lick-index
profiles. Four additional moderately inclined S0 galaxies come from our sample
of nearby early-type disk galaxies -- group members whose central parts have been studied
earlier with the Multi-Pupil Fiber Spectrograph of the 6m telescope; the kinematical
and Lick-index data for the central parts of these galaxies have been partly
published in \citet{me2000,ortho,n3169gr}.
The main global parameters of the galaxies studied in this work are given in Table~1.

\begin{table*}
\scriptsize
\caption[ ] {Global parameters of the galaxies}
\begin{flushleft}
\begin{tabular}{lcrrccccccccc}
\hline\noalign{\smallskip}
Name & Type  & $R^{\prime \prime}_{25}$ & $R_{25}$, kpc  &
$B_T^0$  & $M_B$ & $M_K^3$ & $(B-V)_T^0$  &
$V_r $, $\mbox{km} \cdot \mbox{s}^{-1}$ & D$^4$, Mpc  &
$i_{phot}$  & {\it PA}$_{phot}$  & Environment$^6$ \\
 & (NED$^1$) & (RC3$^2$) &  & (RC3) &  &  & (RC3) & (NED) &
& (LEDA$^5$) & (RC3) & \\
\hline
N502 & SA0$^0$(r) & 34 & 4.9 & 13.57 & --18.8 & --22.6 & 0.95 & 2489 & 30 &
$24^{\circ}$ & -- & 1 \\
N524 & SA0$^+$(rs) & 85 & 11.5 & 11.17 & --21.1 & --25.1 & 1.05 & 2379 & 28 &
$6^{\circ}$ & -- & 2 \\
N1029 & S0/a & 41 & 9.3 & 13.32 & --20.0 & --23.7 & -- & 3635 & 46.5 & $90^{\circ}$
& $70^{\circ}$ & 3 \\
N1032 & S0/a & 97 & 15.7 & 12.29 & --20.4 & --24.3 & 1.00 & 2694 & 34 & $90^{\circ}$ &
$68^{\circ}$ & 5 \\
N1184 & S0/a & 85 & 12.5 & 13.44: & --19.0 & --24.3 & -- & 2342 & 31 & $90^{\circ}$ &
$168^{\circ}$ & 5 \\
N2549 & SA0$^0$(r) & 117 & 8.9 & 12.00 & --19.0 & --22.9 & 0.93 & 1039 & 15.7 & $90^{\circ}$
& $177^{\circ}$ & 4 \\
N2732 & S0 & 63 & 8.2 & 12.85 & --19.3 & --23.2 & 0.96 & 1960 & 27 & $90^{\circ}$ &
$67^{\circ}$ & 4 \\
N3166 & SAB(rs)0/a & 144 & 16.2 & 11.01 & --20.8 & --24.6 & 0.87 & 1345 & 23 & $56^{\circ}$ &
$87^{\circ}$ & 4 \\
N3414 & S0pec & 106 & 12.0 & 11.86 & --20.0 & --23.9 & 0.97 & 1414 & 23.5 & $20^{\circ}$$^7$ &
-- & 2 \\
N4111 & SA0$^+$(r): & 137 & 9.5 & 11.60 & --19.2 & --23.2 & 0.89 & 807 & 14 & $84^{\circ}$ &
$150^{\circ}$ & 0 \\
N4570 & S0 & 114 & 15.5 & 11.80 & --20.5 & --24.55 & 0.94 & 1730 & 28 & $90^{\circ}$ &
$159^{\circ}$ & 0 \\
N5308 & S0$^-$ & 112 & 15.7 & 12.42 & --19.9 & --23.95 & 0.92 & 2041 & 29 & $90^{\circ}$ &
$60^{\circ}$ & 1 \\
N5353 & S0 & 66 & 10.8 & 11.98 & --20.7 & --25.0 & 0.97 & 2325 & 34 & $82^{\circ}$ &
$145^{\circ}$ & 2 \\
N7332 & S0 pec & 122 & 6.6(13.6) & 11.93 & --18.3 (--19.9) & --22.2 (--23.8) & 0.91 & 1172 &
11 (23) & $90^{\circ}$ & $155^{\circ}$ & 4 \\
I1541 & -- & 23$^5$ & 8.5$^5$ & 15.15$^5$ & --19.5$^5$ & --23.5 & 1.04$^8$
& 5926 & 76 & $90^{\circ}$ & $36^{\circ}$ & 1 \\
\hline
\multicolumn{13}{l}{$^1$\rule{0pt}{11pt}\footnotesize
NASA/IPAC Extragalactic Database}\\
\multicolumn{13}{l}{$^2$\rule{0pt}{11pt}\footnotesize
Third Reference Catalogue of Bright Galaxies}\\
\multicolumn{13}{l}{$^3$\rule{0pt}{11pt}\footnotesize
$K_{s,tot}$, from 2MASS, are taken from the NED photometry lists}\\
\multicolumn{13}{l}{$^4$\rule{0pt}{11pt}\footnotesize
Distances from NED, `Cosmology corrected' option, except NGC 7332 where in parentheses
there is $D$ from \citet{sbfdist} }\\
\multicolumn{13}{l}{$^5$\rule{0pt}{11pt}\footnotesize
Lyon-Meudon Extragalactic Database}\\
\multicolumn{13}{l}{$^6$\rule{0pt}{11pt}\footnotesize
Environments: 0 -- cluster member, 1  -- rich-group member,
2 -- rich-group center, 3 -- loose-group member, 4 -- loose-group center,
5 -- field}\\
\multicolumn{13}{l}{$^7$\rule{0pt}{11pt}\footnotesize
An obviously wrong inclination of $77^{\circ}$ is given for NGC~3414
in the Lyon-Meudon Extragalactic Database}\\
\multicolumn{13}{l}{$^8$\rule{0pt}{11pt}\footnotesize
The colour in the central aperture of $3^{\prime \prime}$, according to
\citet{n80ph} }
\end{tabular}
\end{flushleft}
\end{table*}

The sample is small, however the galaxies are homogeneously distributed over the luminosities,
with their blue absolute magnitudes from --19 to --21, and located at different environments.
We have one galaxy (NGC~4570) in the Virgo cluster where the intracluster medium
influence is inavoidable, and one galaxy (NGC~4111) in the Ursa Major cluster where X-ray gas
is not detected. Among group galaxies, NGC~524 and NGC~5353 are central galaxies
embedded into X-ray haloes, NGC~5308 is a member galaxy in the X-ray bright group, and NGC~502
and IC~1541 though being members of rich galaxy groups, lie outside the X-ray
halos of their groups (\citet{xray}, and also some archive ASCA images).
NGC~3414 is a central galaxy in the rich group undetected in X-ray. NGC~2732 is a modest host
of a few faint satellites. NGC~1029, NGC~2549, and NGC~7332 are in triplets.
By using the NED environment searcher, we did not find any galaxies
within 300 kpc from NGC~1032 and NGC~1184 so in our Table~1 we characterized
them as isolated field galaxies.

\section{Observations and data reduction}

The long-slit spectral observations were made with the focal
reducer SCORPIO\footnote{For a description of the SCORPIO instrument,
see http://www.sao.ru/hq/moisav/scorpio/scorpio.html.}
\citep{scorpioman} installed at the prime focus of the Russian 6m
telescope (in the Special Astrophysical Observatory of the Russian Academy of
Sciences). As the main goals were stellar kinematics and Lick indices H$\beta$,
Mgb, Fe5270, and Fe 5335, we observed a quite narrow spectral range rich with
absorpion lines, 4800--5500~\AA, by using the volume-phase grating 2300G. The slit
width was one arcsecond and the spectral resolution -- about 2~\AA.
The CCD $2k \times 2k$ EEV CCD42-40 and later in 2010 -- the CCD $2k \times 4k$
E2V CCD42-90 were used as detectors, and the sampling along the slit was
$0.36\arcsec$ per pixel. The slit length is about 6 arcminutes so the data from
the edges of the slit were used as the sky background to be subtracted from the
galaxy spectra. Inhomogeneties of the optics transmission and spectral resolution
along the slit were checked with the high signal-to-noise twilight exposures.
Since some of lenticular galaxies reveal weak emission lines in their spectra,
and their stellar Lick index H$\beta$ may be contaminated by the Balmer emission
line of the ionized gas, we also obtained spectra of a redder spectral range for these galaxies.
This was done using the volume-phase grating 1800R (6100--7100~\AA), providing the
spectral resolution of $\sim 3$~\AA, or the the grating 1200R (5700--7400~\AA),
providing the spectral resolution of $\sim 5$~\AA. These data have been used to calculate the equivalent width
of the H$\alpha$ emission line. In order to do this for the bulge-dominated area,
we summed the spectra over $1\arcsec -3\arcsec$ intervals near the radius
of $0.5r_e$ for every bulge, and then made Gaussian multi-component fitting of
the [NII]$\lambda$6548+6583+H$\alpha$(emission)+H$\alpha$(absorption) line blend.
The derived
equivalent widths of the H$\alpha$ emission line were used to calculate the correction
for the H$\beta$ index as it was described by \citet{me2006}. The disks in our sample 
are mostly emission-free.The journal of all long-slit observations is presented in Table~2.

\begin{table*}
\caption[ ] {Long-slit spectroscopy of the sample galaxies}
\begin{flushleft}
\begin{tabular}{lrrrcl}
\hline\noalign{\smallskip}
Galaxy & Date & Exposure & PA(slit)
& Spectral range & Seeing \\
\hline\noalign{\smallskip}
NGC~502 & 03 Sep 08 & 80 min &
 $65^{\circ}$ & 4800-5500~\AA\ & $2^{\prime \prime}$ \\
NGC~502 & 03 Sep 08 & 80 min &
 $155^{\circ}$ & 4800-5500~\AA\ & $2^{\prime \prime}$ \\
NGC~524 & 17 Aug 07 & 40 min &
$23^{\circ}$ & 6100-7100~\AA\ & $1\farcs 5$ \\
NGC~524 & 19 Oct 07 & 160 min &
$127^{\circ}$ & 4800-5500~\AA\ & $2^{\prime \prime}$ \\
NGC~524 & 04 Sep 08 & 45 min &
$38^{\circ}$ & 6100-7100~\AA\ & $1\farcs 3$ \\
NGC~524 & 04 Sep 08 & 20 min &
$115^{\circ}$ & 6100-7100~\AA\ & $1\farcs 3$ \\
NGC~1029 & 6 Nov 10 & 90 min &
$70^{\circ}$ & 4650-5730~\AA\ & $2\farcs 5$ \\
NGC~1032 & 16 Oct 09 & 100 min &
$68^{\circ}$ & 6100-7100~\AA\ & $2\farcs 1$ \\
NGC~1032 & 17 Oct 09 & 120 min &
$68^{\circ}$ & 4825-5500~\AA\ & $2\farcs 9$ \\
NGC~1184 & 14 Oct 09 & 120 min &
$168^{\circ}$ & 4825-5500~\AA\ & $2\farcs 9$ \\
NGC~1184 & 15 Oct 09 & 180 min &
$168^{\circ}$ & 4825-5500~\AA\ & $2\farcs 3$ \\
NGC~2549 & 7 Dec 10 & 180 min &
$177^{\circ}$ & 4650-5730~\AA\ & $2\farcs 5$ \\
NGC~2549 & 25 Dec 10 & 150 min &
$145^{\circ}$ & 5700-7400~\AA\ & $3^{\prime \prime}$ \\
NGC~2732 & 12 Oct 09 & 120 min &
$152^{\circ}$ & 6100-7100~\AA\ & $3^{\prime \prime}$ \\
NGC~2732 & 15 Oct 09 & 120 min &
$67^{\circ}$ & 4825-5500~\AA\ & $2\farcs 8$ \\
NGC~2732 & 15 Oct 09 & 120 min &
$67^{\circ}$ & 6100-7100~\AA\ & $1\farcs 7$ \\
NGC~3166 & 28 Apr 06 & 80 min &
$86^{\circ}$ & 4800-5500~\AA\ & $2\farcs 8$ \\
NGC~3414 & 30 Mar 09 & 45 min &
$20^{\circ}$ & 6100-7100~\AA\ & $1\farcs 5$ \\
NGC~3414 & 31 Mar 09 & 100 min &
$150^{\circ}$ & 6100-7100~\AA\ & $1\farcs 3$ \\
NGC~3414 & 11 Apr 10 & 140 min &
$150^{\circ}$ & 4650-5730~\AA\ & $3^{\prime \prime}$ \\
NGC~4111 & 30 Mar 09 & 45 min &
$150^{\circ}$ & 6100-7100~\AA\ & $2^{\prime \prime}$ \\
NGC~4111 & 21 May 09 & 140 min &
$150^{\circ}$ & 4825-5500~\AA\ & $1\farcs 3$ \\
NGC~4111 & 19 Dec 09 & 105 min &
$240^{\circ}$ & 4825-5500~\AA\ & $3\farcs 6$ \\
NGC~4570 & 08 Apr 10 & 120 min &
$159^{\circ}$ & 4650-5730~\AA\ & $3^{\prime \prime}$ \\
NGC~5308 & 10 Apr 10 & 100 min &
$60^{\circ}$ & 4650-5730~\AA\ & $3^{\prime \prime}$ \\
NGC~5353 & 05 Apr 09 & 45 min &
$145^{\circ}$ & 6100-7100~\AA\ & $1\farcs 7$ \\
NGC~5353 & 24 May 09 & 60 min &
$145^{\circ}$ & 4825-5500~\AA\ & $3\farcs 4$ \\
NGC~7332 & 11 Oct 09 & 180 min &
$155^{\circ}$ & 4825-5500~\AA\ & $1\farcs 1$ \\
NGC~7332 & 11 Oct 09 & 100 min &
$155^{\circ}$ & 6100-7100~\AA\ & $2\farcs 5$ \\
IC~1541 & 2 Sep 08 & 30 min &
$25^{\circ}$ & 5700-7400~\AA\ & $2\farcs 5$ \\
IC~1541 & 2 Nov 10 & 90 min &
$35^{\circ}$ & 4650-5730~\AA\ & $1\farcs 2$ \\
\hline
\end{tabular}
\end{flushleft}
\end{table*}

\section{Photometric structure of the sample galaxies}

In order to match the stellar population properties to particular galaxy components, 
such as disks and bulges,
we have first of all to analyse the light distribution in these galaxies.

The structural parameters of the edge-on sample galaxies are mostly taken
from the work by \citet{edgeonphot} supplemented by \citet{dopphot}
where $K_s$-band 2MASS images are decomposed
by means of the BUDDA software \citep{budda}: these galaxies are NGC~1029, 1184,
2549, 2732, 4111, 5308, 4570, 5353, and 7332. Since we know the 2MASS images to 
be rather shallow, we have checked the validity of the disk exponential scale lengths 
by decomposing the latest SDSS data for the most of these galaxies in the $r$-band 
with the GALFIT \citep{galfit}; the discrepancy of the found $r$-band scale lengths
with those in the $K_s$-band is of order of a few arcseconds, or within
the $K_s$-band scale length estimate accuracy. The bulge effective radius
for NGC~1032 is taken from \citet{gorgasprof}.

Detailed study of the photometric structures of S0 galaxies IC~1541, NGC~502,
NGC~524, and NGC~2732 has been undertaken by us through the $BV$- and
$R-$imaging with the focal reducer SCORPIO of the 6m
telescope operating in imaging mode. The photometric
structure of IC~1541 in the $BV$-bands has been published by us
earlier \citep{n80ph}. In face-on NGC~502 and NGC~524 of the NGC~524 group
we have found compact bulges and two-tiered (antitruncated) large-scale
exponential disks starting to dominate at $R\approx 15\arcsec -20\arcsec$ \citep{n524gr}.

NGC~3166 is a luminous early-type disk galaxy and the member of a loose group where
it is located near the center being almost as luminous as its nearest neighbour, NGC~3169.
The structure of the galaxy is very complex, particularly in its central part. Although
the galaxy is inclined to our line of sight by at least $60^{\circ}$, with its line of
nodes oriented nearly E--W, the ellipticity of the isophotes drops to zero at
the radius of $\sim 15\arcsec -20\arcsec$. \citet{osu}, using the results of their OSU
Nearby Galaxies Survey, have identified the low-surface brightness bar aligned perpendicular
to the line of nodes. There were efforts to
decompose the photometric structure of the galaxy: \citet{fisherdrory} analysed the
surface brightness profile in the radius range of $2\arcsec -45\arcsec$ and found
only a pseudobulge with the Sersic index of $n=0.56$,
and \citet{lauri10} found three (!) bars in the central part of the galaxy
though failing to fit the outer disk. We have adopted the bulge effective radius
$r_e=5\farcs 4$ in the $K_s$-band from \citet{lauri10}; however, since the galaxy
possesses a compact edge-on circumnuclear disk well seen up to $R=5\arcsec -6\arcsec$,
for this galaxy we estimate the bulge stellar population properties {\it beyond}
this circumnuclear disk, at $R=7\farcs 5 - 9\arcsec$, and not at $R=0.5r_e$.
By decomposing the SDSS $r$-band image, we have succeeded to derive parameters
of a very extended, low-surface brightness large-scale disk of the galaxy; it is
the only example of our sample where we have not reached the radius of $2h$ with
our spectral data because its $h>5$~kpc.

The bulge effective radii in the $K_s$-band for NGC~524 and NGC~3414 are
taken from \citet{lauri10}.

The decomposition results in the optical bands for the moderately inclined sample galaxy
NGC~3414 are presented below.

\subsection{The global structure of NGC~3414}

Due to its peculiar appearance and strong non-axisymmetry, the structure of  NGC~3414 requires 
a careful approach. There are controversial points of view on the 3D orientation of this galaxy:
\citet{bag} and \citet{chitrejog} consider it
as a face-on S0 with a high-contrast thin bar while \citet{polarcat}
and \citet{lauri05} treat it as an edge-on galaxy with a large-scale
polar disk.

\begin{figure*}
\hfil
\begin{tabular}{c c}
 \includegraphics[width=8cm]{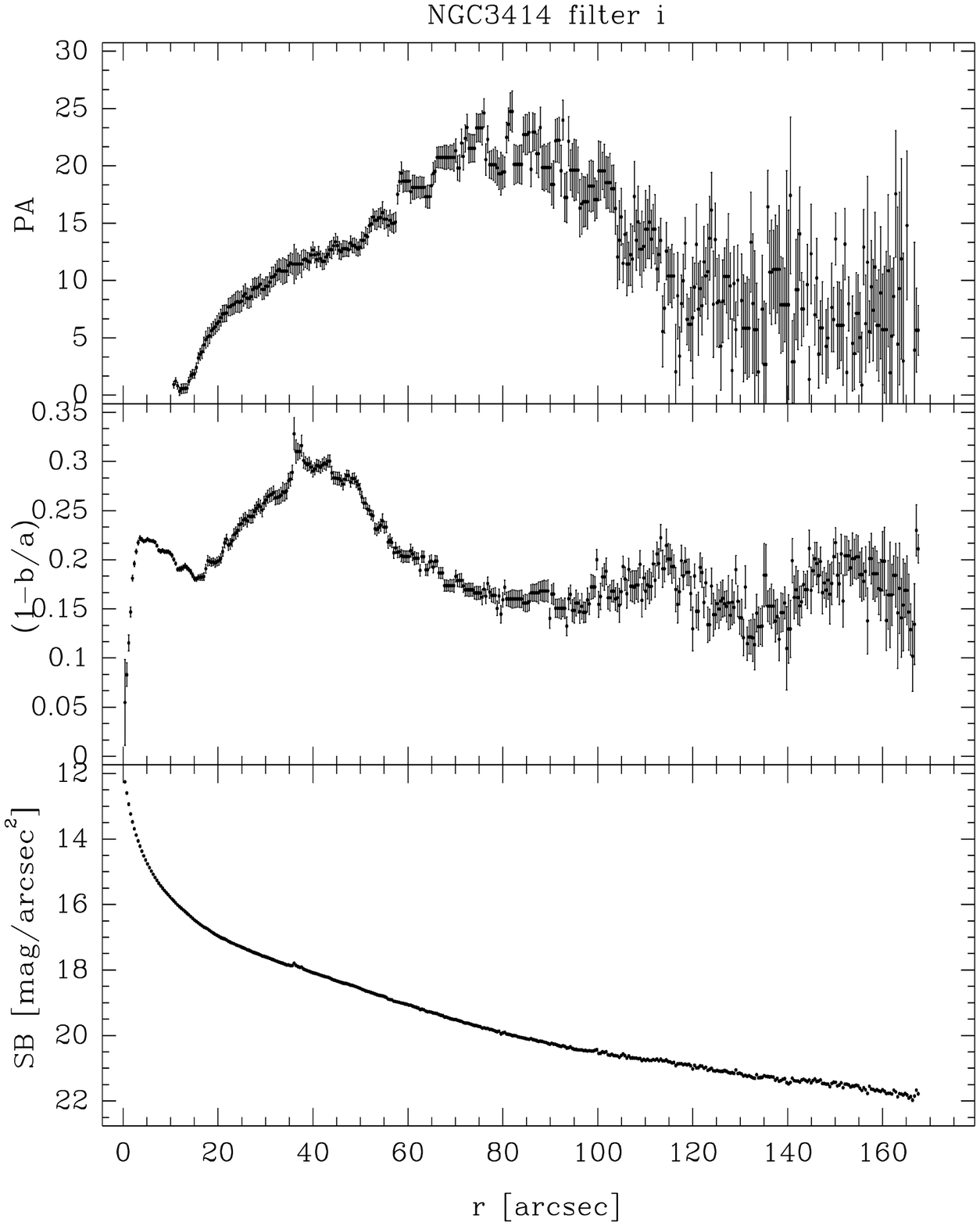} &
 \includegraphics[width=8cm]{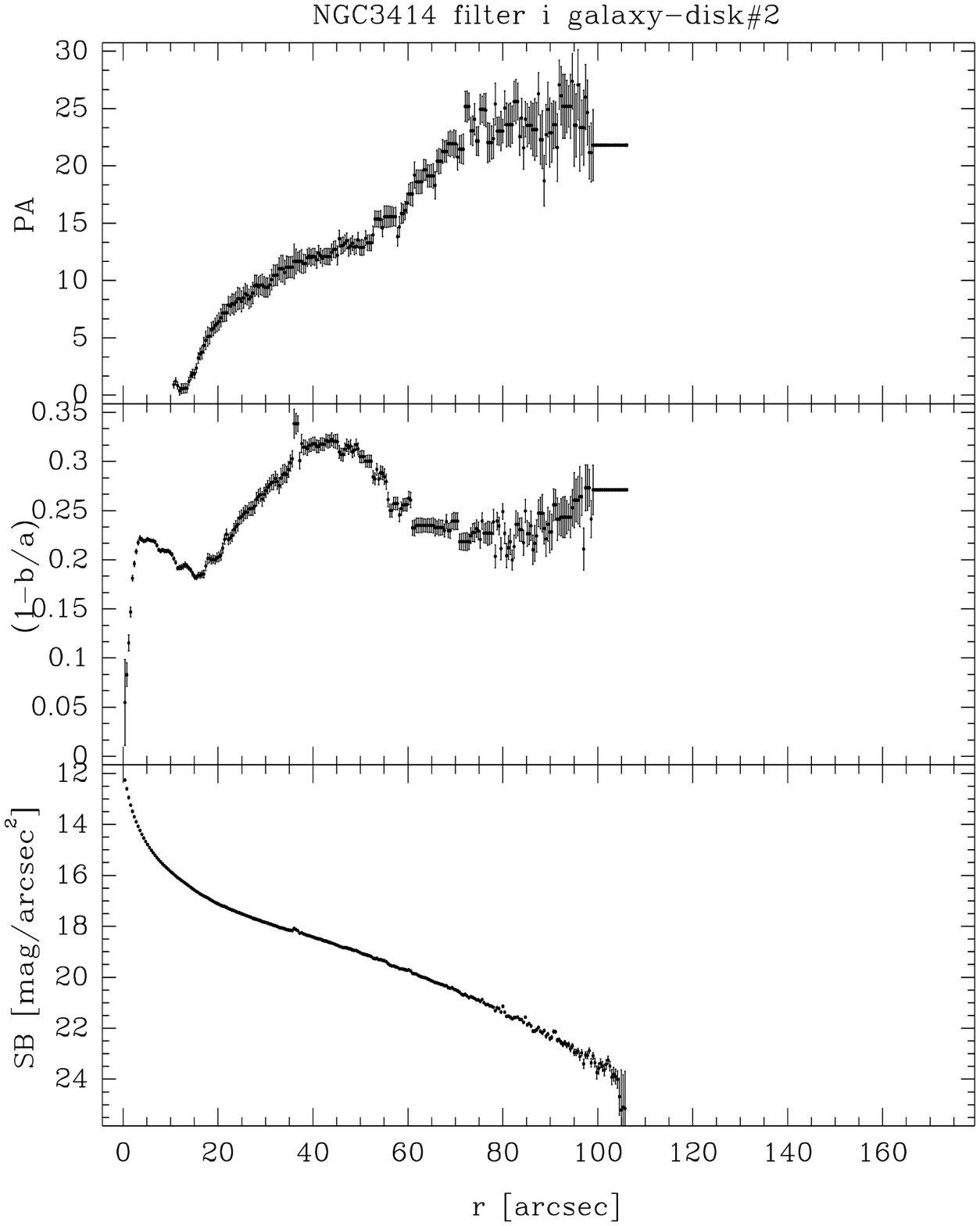} \\
 (a)&(b)\\
\end{tabular}
\caption{The results of the isophotal analysis
for the $i$-band image of NGC~3414: {\it left} -- for the full image,
{\it right} -- for the residual image after subtracting the model outer disk. In both plots
{\it top} -- the position angle of the isophote major axis, {\it middle} --
the isophote ellipticity, {\it bottom} -- the azimuthally averaged surface brightness
profiles.}
\label{n3414prof}
\end{figure*}

We analysed a photometric structure of the galaxy by applying
the 2D decomposition software BUDDA \citep{budda} to the data taken
from the SDSS/DR7 archive. The images were sky-subtracted according to the header
notification.  The used version of the BUDDA allowed to decompose a galaxy
into one exponential disk and a Sersic bulge. However we know that the slope of
the exponential disk profile may change along the radius -- disks may be two-tiered,
truncated or antitruncated \citep{pt06,erwinbars}.
To take into account this possibility, we have invented a more complex
approach to the galaxy decomposition. Basing on isophotal analysis results,
we determine an (outer) radius range where the thin disk dominates -- it is
the (outer) zone with the isophote ellipticity staying constant along the radius at
a value corresponding to the $1 - \cos i$ where $i$ is the disk inclination
to the line of sight. The isophote analysis has been made in the frame of the IRAF
software (Fig.~\ref{n3414prof}).
After that we masked the inner part of the galaxy and applied the BUDDA
procedure only to the outer disk-dominated zone. After obtaining the parameters
of the outer disk, we constructed a 2D model disk image, subtracted it from the
full galaxy image and then applied the BUDDA for the second time, this time to the
residual image, to derive the parameters of the inner disk and of the bulge.
In NGC~3414 the isophote ellipticity comes to a constant level only at $R>85\arcsec$,
and it is a rather low ellipticity: the outer stellar disk is indeed close to face-on
orientation. The inner disk, inside $R=80\arcsec$, has a shorter scale length,
a higher visible ellipticity, and contains perhaps a bar (Fig.~\ref{n3414prof}).
The full decomposition results in three bands, $gri$, are given in the Table~3.

\begin{table*}
\scriptsize
\caption[ ] {NGC~3414: Parameters of the brightness profile fitting by a three-component
 Sersic model}
\begin{flushleft}
\begin{tabular}{|l|ccccccc|}
\hline
Component & Radius range of fitting, arcsec & n
 & $PA_0$  & $1-b/a$ &
$h_0$, arcsec & $h_0$, kpc & $r_e$, arcsec \\
\hline
\multicolumn{8}{l}{NGC 3414, $i$-band}\\
Outer disk & $>85$ & 1 & $3\degr$ & $0.05\pm 0.03$ &
 $56 \pm 14$ & $6.4 \pm 1.6 $ & -- \\
Inner disk & 20--70 & 1 & $25\degr \pm 4\degr$ & $0.27 \pm 0.05$ &
 $18 \pm 2$ & $2.0 \pm 0.2$ & -- \\
Central bulge & $<15$ & $2.1\pm 0.4$ &  $10\degr \pm 8\degr$ & $0.18 \pm 0.05$
& -- & -- & $3.8 \pm 0.4$ \\
\multicolumn{8}{l}{NGC 3414, $r$-band}\\
Outer disk & $>85$ & 1 & $3\degr$ & $0.08\pm 0.05$ &
  $63 \pm 19$ &  $7.2 \pm 2.2$ & -- \\
Inner disk & 20--70 & 1 & $25\degr \pm 4\degr $ & $0.26 \pm 0.05$ &
 $17 \pm 2$ & $1.9 \pm 0.2$  & -- \\
Central bulge & $<15$ & $2.1\pm 0.4$ &  $10\degr \pm 7\degr $ & $0.18 \pm 0.05$ &
-- & -- & $3.5 \pm 0.3$ \\
\multicolumn{8}{l}{NGC 3414, $g$-band}\\
Outer disk & $>85$ & 1 & $3\degr$ & $0.06\pm 0.04$ &
  $48 \pm 16$ &  $5.5 \pm 1.8$ &  -- \\
Inner disk & 20--70 & 1 & $25\degr \pm 5\degr $ & $0.31 \pm 0.07$ &
 $17 \pm 2$ & $1.9 \pm 0.2$ & -- \\
Central bulge & $<15$ & $2.2\pm 0.9$ &  $10\degr \pm 11\degr $ & $0.15 \pm 0.05$ &
-- & -- & $3.0 \pm 0.4$ \\
\hline
\end{tabular}
\end{flushleft}
\end{table*}

\section{Age and metallicity along the radius}

Focussing on the H$\beta$, Mgb, Fe5270, Fe5335, we have measured the Lick index profiles up to
large distances from the galaxy centers exceeding two exponential scale lengths in almost all cases.
The Lick index system was controlled for every observational run with a sample of standard
Lick stars \citep{woretal} in the way described by \citet{webaes}.

\begin{figure*}
\hfil
\begin{tabular}{c c}
 \includegraphics[width=8cm]{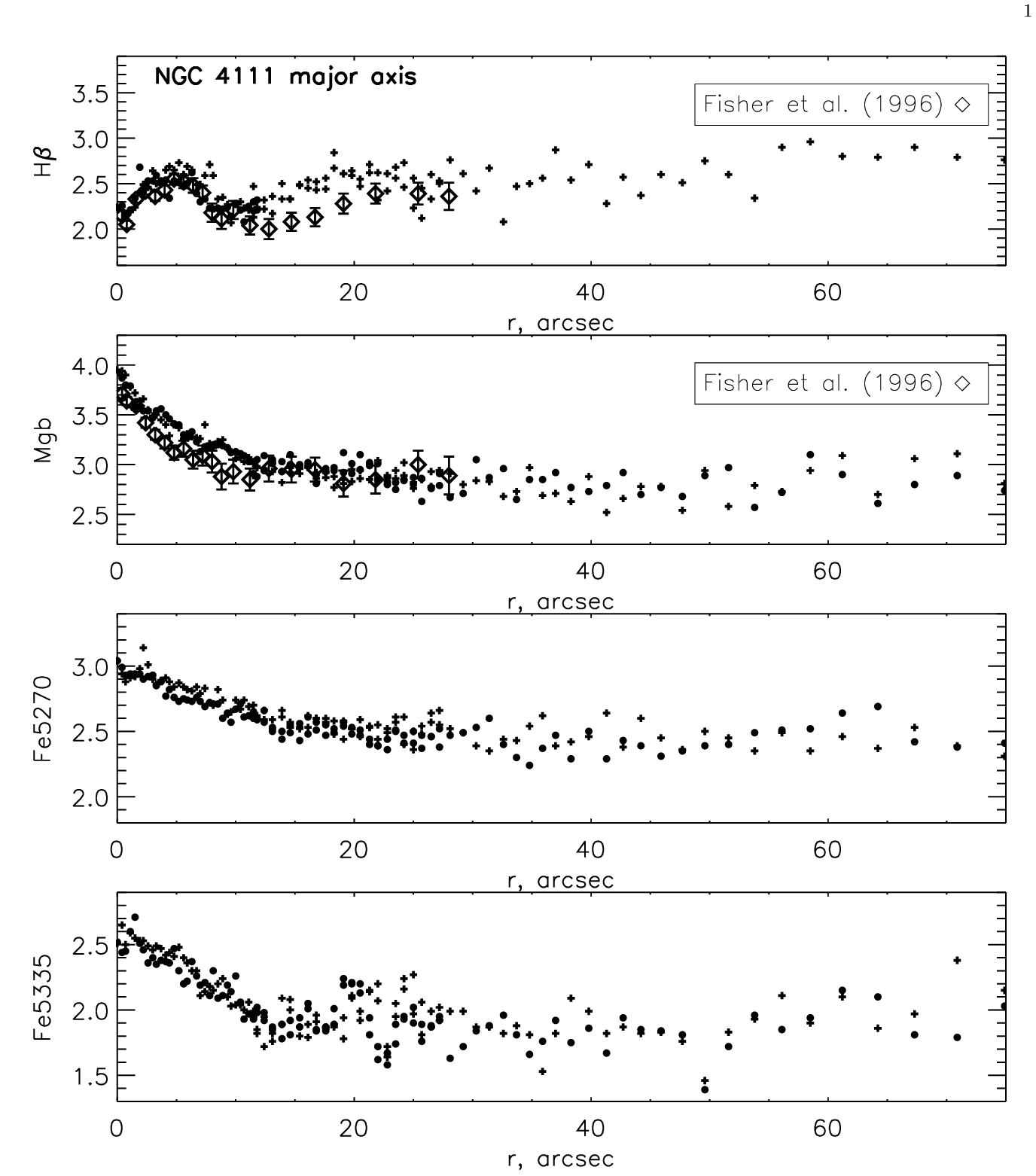} &
 \includegraphics[width=8cm]{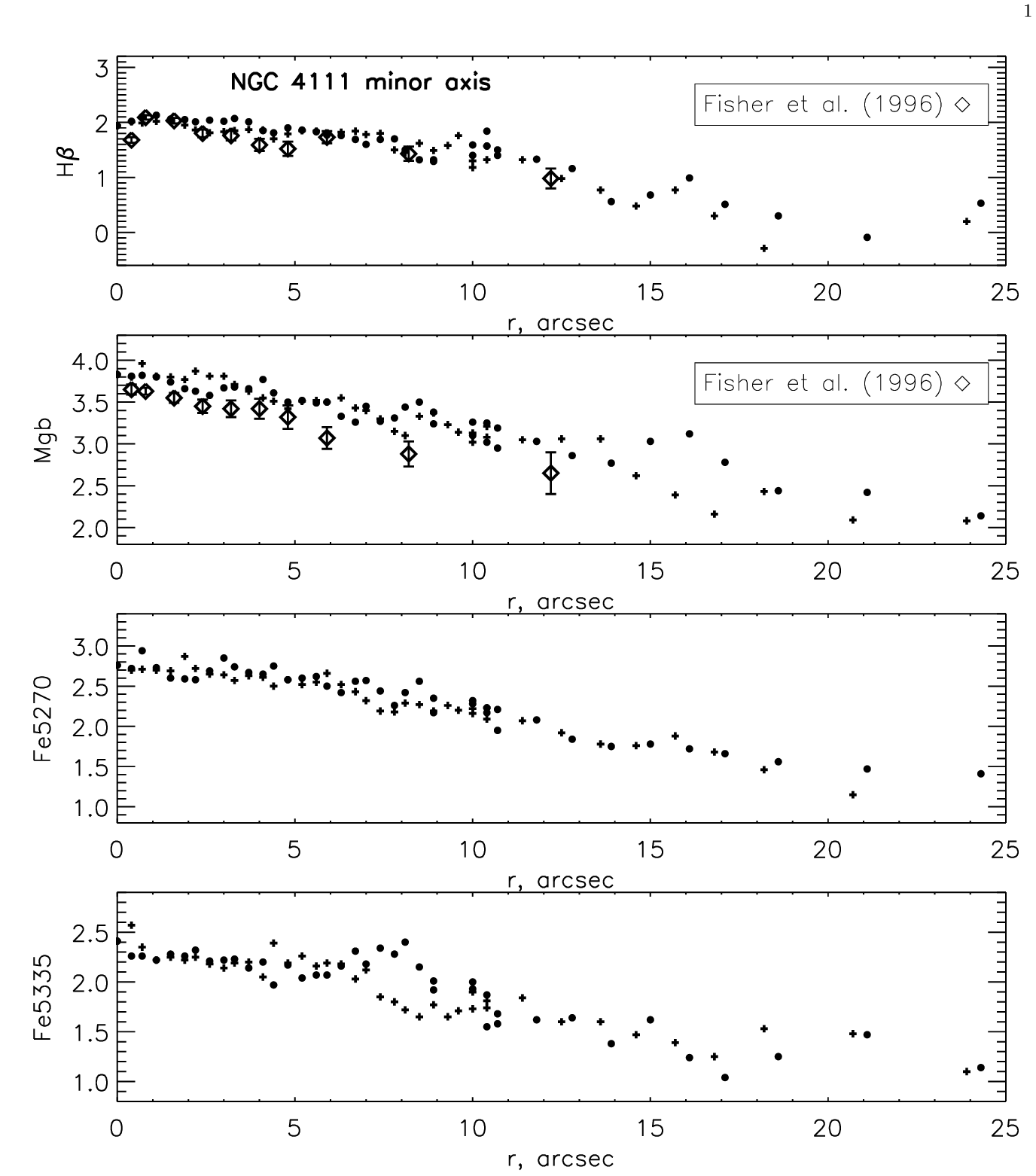} \\
 (a)&(b)\\
\end{tabular}
\caption{Radial profiles of the Lick indices H$\beta$, Mgb, Fe5270, and Fe5335 as
measured from the SCORPIO spectral data along the major ({\it a}) and minor ({\it b})
axes of NGC~4111; dots and crosses refer to different sides of the slit with respect
to the center, the data by \citet{fish96} are overplotted for comparison.}
\label{n4111ind}
\end{figure*}

In Fig.~\ref{n4111ind} we show the Lick index profiles measured along the major and 
minor axes in NGC~4111. The comparison to the previous measurements of Mgb and H$\beta$ 
along the major and minor axes done for this galaxy by \citet{fish96} shows that our profiles 
are twice as extended. Two halves of the profiles, the northern and southern ones, are in good agreement with each other, with the exception of the H$\beta$
profile where we have not been able to measure the northern part because of the low recession
velocity of the stellar component resulting in a cut off the blue-continuum band of the
H$\beta$ index. The low point-to-point scatter till the last measured radii gives
evidence for the high accuracy of the Lick index measurements even in the outer part
of the disk.

\begin{figure*}
\hfil
\begin{tabular}{c c}
 \includegraphics[width=8cm]{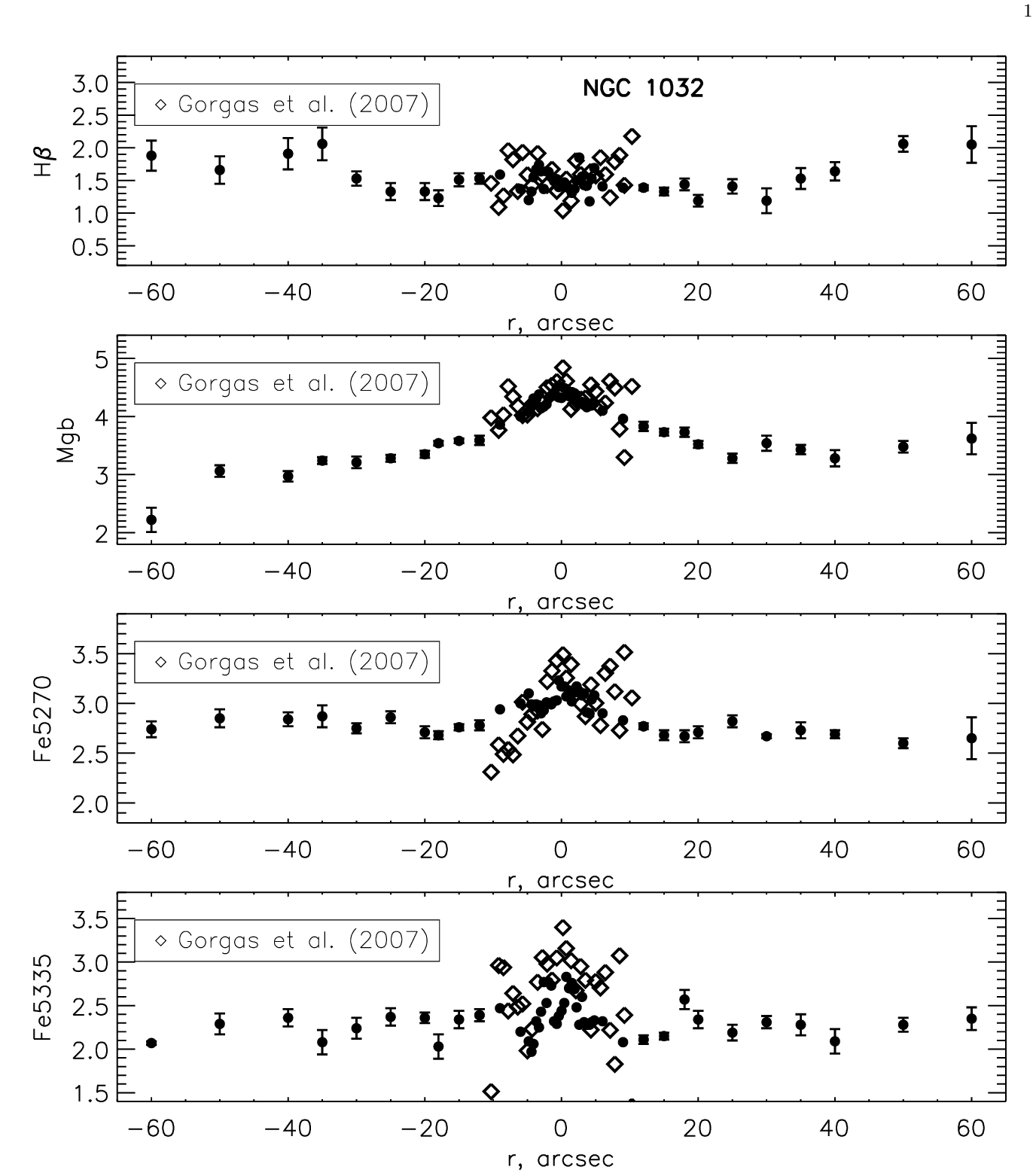} &
 \includegraphics[width=8cm]{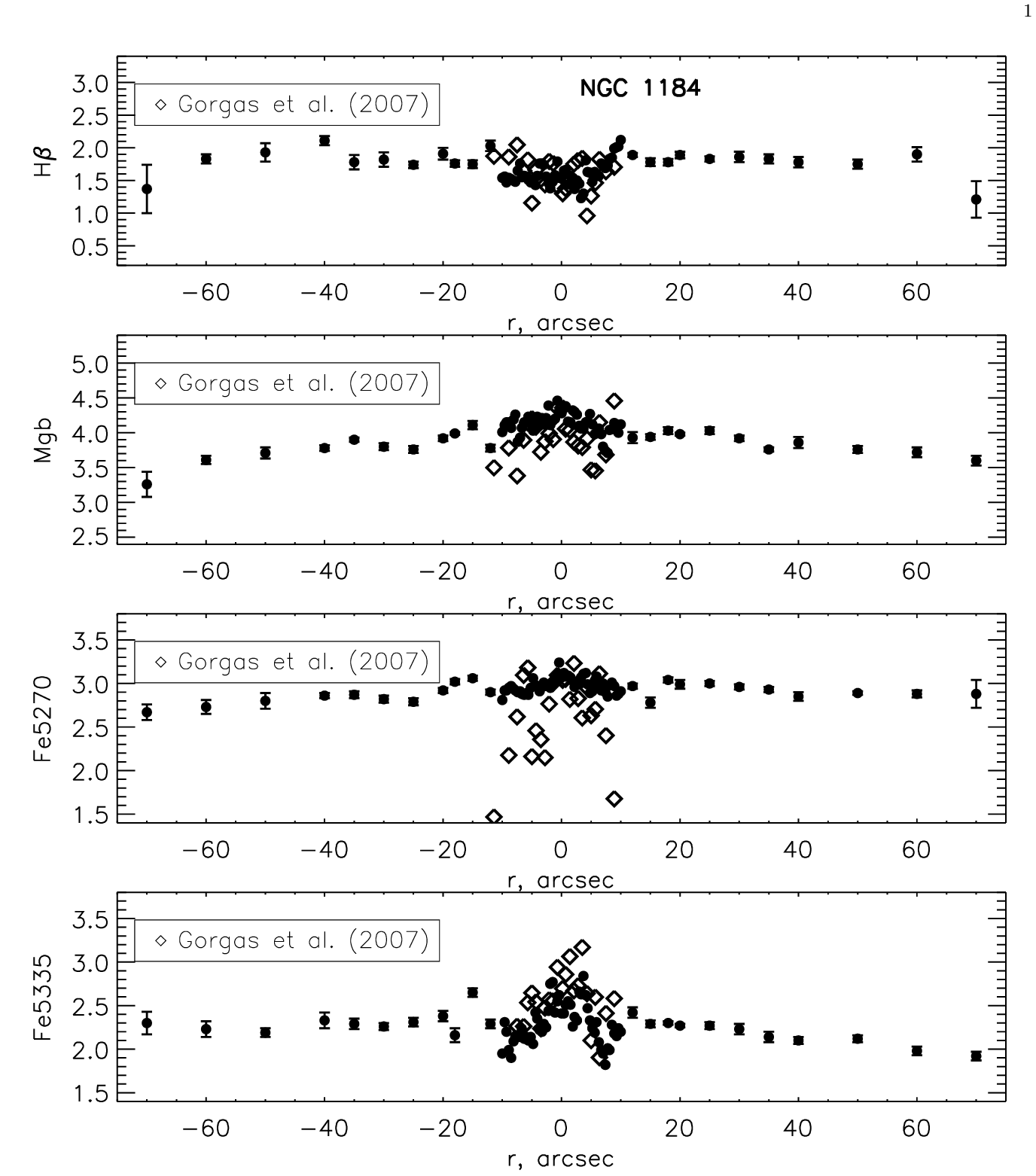} \\
(a) & (b) \\
\includegraphics[width=8cm]{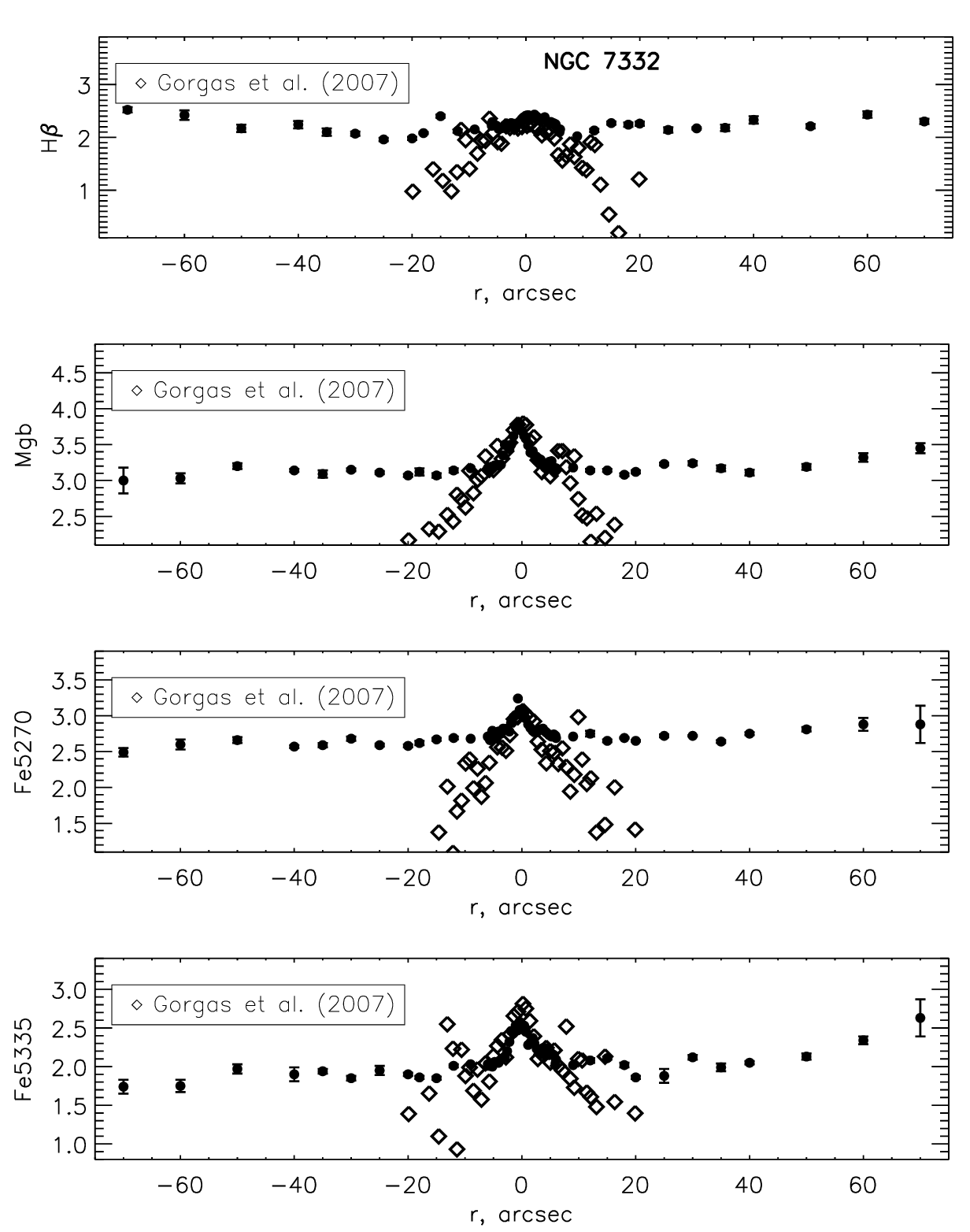} &
\includegraphics[width=8cm]{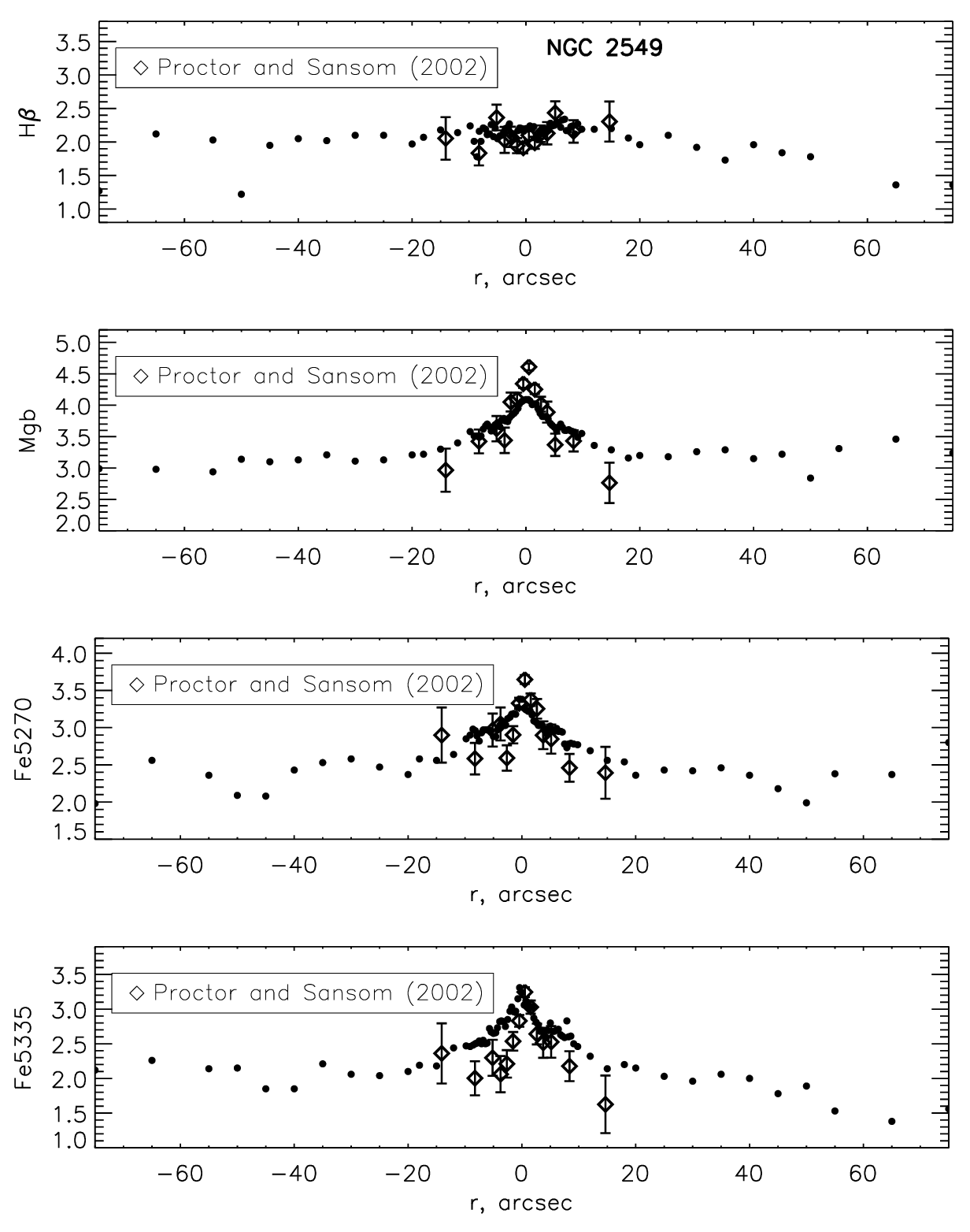} \\
 (c) &  (d)\\
\end{tabular}
\caption{Radial profiles of the Lick indices H$\beta$, Mgb, Fe5270, and Fe5335 as
measured from the SCORPIO spectral data in ({\it a}) NGC~1032, ({\it b}) NGC~1184,
({\it c}) NGC~7332, and ({\it d}) NGC~2549 along their major axes. For comparison, the analogous
profiles along the {\it minor} axes in the same galaxies published by \citet{gorgasprof}
and \citet{proctor} are overplotted.}
\label{otherind}
\end{figure*}

Figure~\ref{otherind} presents four more examples of our Lick index profiles 
measured along the major axes of the edge-on lenticular galaxies. One can see very
compact central parts with rapid variations of the Lick indices; these are the bulge-dominated
zones. Outside these zones, the Lick index profiles look rather flat. For the outer parts, we
have binned index measurements in radial ranges of $5\arcsec -10\arcsec$; the errors shown
are standard errors of the means. As a comparison to our data, in Fig.~\ref{otherind} we have also
plotted the results of \citet{gorgasprof} and \citet{proctor} obtained from long-slit spectra along
the {\it minor} axes of the galaxies. Their studies were concentrated exclusively on the bulges.
In the centers of the profiles the agreement with our data is almost perfect which confirms
our good calibration onto the standard Lick index system and insignificant contributions
of the disks into the spectra along the major axes at $R<5\arcsec - 10\arcsec$.

Let us consider separately the bulges and the disks of the galaxies in our total sample.

\subsection{The bulges}

In order to characterize the bulges, we have decided to probe their stellar population properties at
the radii of $0.5r_e$ where $r_e$ refer the effective radii of the bulges {\it after} the surface brightness profile decomposition. Since we have photometric decomposition results for all our galaxies,
we have estimated the contribution of the disks to the surface brightness at $R=0.5r_e$ by extrapolating
the outer exponential disk profiles toward the centers of the galaxies. These estimates can be used to correct
the bulge Lick indices measured at $R=0.5r_e$ for the disk contributions by assuming a constant level
of disk Lick indices over the whole galaxies. In all cases, the disk contributions are rather
small and affect the bulge Lick indices only within 0.15~\AA\ for H$\beta$, 0.4~\AA\ for Mgb, and
0.2~\AA\ for $\langle \mbox{Fe} \rangle \equiv (\mbox{Fe5270} + \mbox{Fe5335})/2$.
The halves of the effective radius values, mostly in the $K_s$-band, and the corrected 
Lick indices H$\beta$, Mgb, and $\langle \mbox{Fe} \rangle$ for the bulges are listed in
Table~4. Seven galaxies of 15 have noticeable Balmer emission lines in their central spectra; the
equivalent widths of the H$\alpha$ emission lines, which are measured at $R=0.5r_e$ by applying
Gaussian multi-component fitting to the total line profiles  taking into account also the wide stellar H$\alpha$
absorption line, are listed in Table~4 too. Furthermore, before using the H$\beta$ Lick index
to determine the stellar population age and metallicitites, we calculate the H$\beta$ index
corrections for the emission. The H$\beta$ emission-line
intensities are related to the H$\alpha$ emission-line intensities by ionization
models under the assumption of the excitation mechanisms. The largest correction corresponds to
the gas excitation by young stars, $\Delta \mbox{H}\beta \approx 0.4 EW(\mbox{H}\alpha \mbox{emis})$
\citep{burgess}, other mechanisms including effects of dust give steeper Balmer decrements.
Our galaxies are of early type and do not reveal clear signs of current star formation.
To calculate the H$\beta$ corrections, we choose the empirical Balmer decrement found by
\citet{sts2001} for a large sample of disk galaxies which corresponds
to the mixed gas excitation nature. The detailed description of the procedure can be
found in \citet{me2006}. In any case, for the galaxies in our sample the largest possible
correction is below 0.4~\AA.

\begin{table*}
\scriptsize
\caption[ ]{Stellar population parameters of the bulges}
\begin{flushleft}
\begin{tabular}{lcrrrrrrr}
\hline\noalign{\smallskip}
NGC/IC & $0.5r_e$ & $EW$(H$\alpha$), \AA & H$\beta$ & Mgb & $\langle \mbox{Fe} \rangle$ &
$T$,~Gyr & $\mbox{[Z/H]}$ & $\mbox{[Mg/Fe]}$ \\
\hline
N502 & $1\farcs 5$ & -- & 1.78 & 4.41 & 2.88 & $6\pm 0.5$ & $+0.3$ & $+0.20$ \\
 & & & $\pm 0.02$ &  $\pm 0.04$ & $\pm 0.02$ & & & $\pm 0.01$ \\
N524 & $4\farcs 4 $ & 0.53 & 1.67 & 4.39 & 2.41 & $11\pm 1$ & $+0.0$ & $+0.35$ \\
 & & & $\pm 0.02$ & $\pm 0.01$ & $\pm 0.04$ & & & $\pm 0.02$ \\
N1029 & $2\arcsec $ & 0.3$^*$ & 1.53 & 4.81 & 3.10 & $10\pm 2$ & $+0.4$ & $+0.18$ \\
 & & & $\pm 0.09$ & $\pm 0.07$ & $\pm 0.11$ & & & $\pm 0.05$ \\
N1032 & $6\arcsec $ & 0.95 & 1.51 & 4.38 & 2.67 & $8\pm 2$ & $+0.2$ & $+0.26$ \\
 & & & $\pm 0.09$ & $\pm 0.07$ & $\pm 0.07$ & & & $\pm 0.04$ \\
N1184 & $6\arcsec $ & -- & 1.43 & 4.23 & 2.57 & $15\pm 2$ & $0.0$ & $+0.25$ \\
 & & & $\pm 0.04$ & $ \pm 0.04$ & $\pm 0.03$ & & & $\pm 0.02$ \\
N2549 & $6\arcsec $ & -- & 2.32 & 3.71 & 2.89 & $2\pm 0.3$ & $+0.5$ & $+0.13$ \\
 & & & $\pm 0.04$ & $\pm 0.02$ & $\pm 0.02$ & & & $\pm 0.01$ \\
N2732 & $4\arcsec $ & 0.38 & 2.09 & 3.75 & 2.58 & $3\pm 0.5$ & $+0.2$ & $+0.22$ \\
 & & & $\pm 0.04$ & $\pm 0.03$ & $ \pm 0.02$ & & & $\pm 0.01$ \\
N3166 & $2\farcs 7$ & -- & 2.16 & 3.26 & 2.64 & $4\pm 0.7$ & $+0.1$ & $+0.08$ \\
 & & & $\pm 0.07$ & $\pm 0.06$ & $\pm 0.07$ & & & $\pm 0.04$ \\
N3414 & $3\arcsec $ & 0.61 & 1.23 & 4.56 & 2.75 & $>12$ & $+0.0$ & $+0.25$ \\
 & & & $\pm 0.07$ & $\pm 0.03$ & $\pm 0.04$ & & & $\pm 0.02$ \\
N4111 & $5\arcsec$ & 0.14(N)--0.97(S) & 2.67 & 3.52 & 2.70 & $<2$ & $>+0.3$ & $+0.16$ \\
 & & & $\pm 0.04$ & $\pm 0.03$ & $\pm 0.02$ & & & $\pm 0.01$ \\
N4570 & $5\farcs 5$ & -- & 1.51 & 4.19 & 2.62 & $14\pm 2$ & $0.0$ & $+0.23$ \\
 & & & $\pm 0.09$ & $\pm 0.04$ & $\pm 0.04$ & & & $\pm 0.02$ \\
N5308 & $5\farcs 5$ & -- & 1.49 & 4.32 & 2.48 & $15\pm 2$ & $-0.05$ & $+0.30$ \\
 & & & $\pm 0.05$ & $\pm 0.05$ & $\pm 0.03$ & & & $\pm 0.02$ \\
N5353 & $6\arcsec $ & -- & 1.63 & 5.05 & 2.98 & $7\pm 2$ & $+0.5$ & $+0.28$ \\
 & & & $\pm 0.08$ & $\pm 0.04$ & $\pm 0.02$ & & & $\pm 0.02$ \\
N7332 & $4\farcs 5$ & 0.22 & 2.28 & 3.22 & 2.45 & $3\pm 0.3$ & $+0.1$ & $+0.16$ \\
 & & & $\pm 0.02$ & $\pm 0.01$ & $\pm 0.02$ & & & $\pm 0.01$ \\
I1541 & $2\farcs 5 $ & -- & 2.14 & 4.09 & 2.50 & $3.3\pm 0.7$ & $+0.25$ & $+0.31$ \\
 & & & $\pm 0.10$ & $\pm 0.05$ & $\pm 0.10$ & & & $\pm 0.05$ \\
\hline
\multicolumn{9}{l}{$^{*}$\rule{0pt}{11pt}\footnotesize
Measured by \citet{ngs_sp} in the central $7^{\prime \prime}
\times 3^{\prime \prime}$ spectrum of NGC 1029.}\\
\end{tabular}
\end{flushleft}
\end{table*}

\begin{figure*}
\hfil
\begin{tabular}{c c}
\includegraphics[width=8cm]{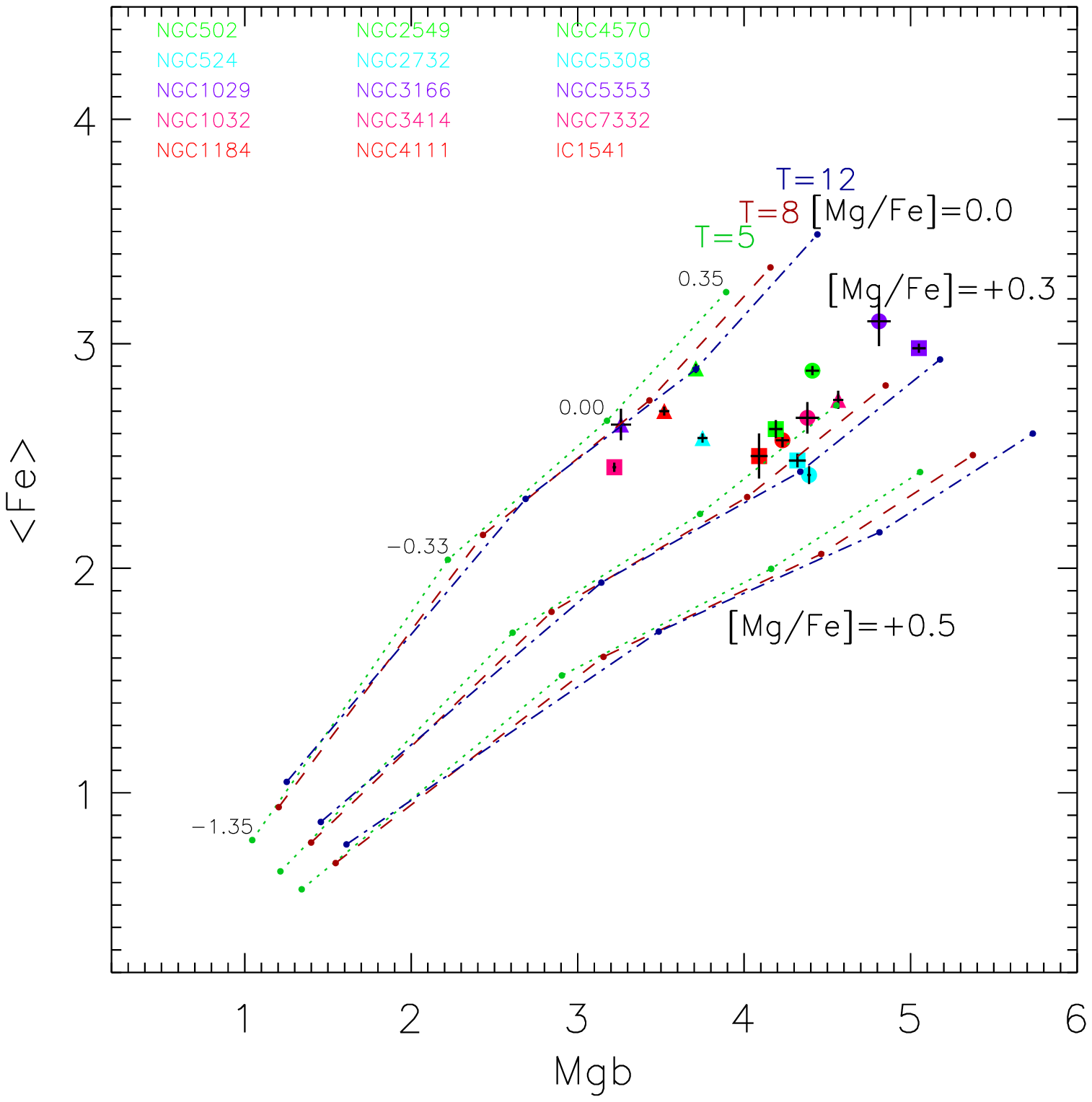} &
 \includegraphics[width=8cm]{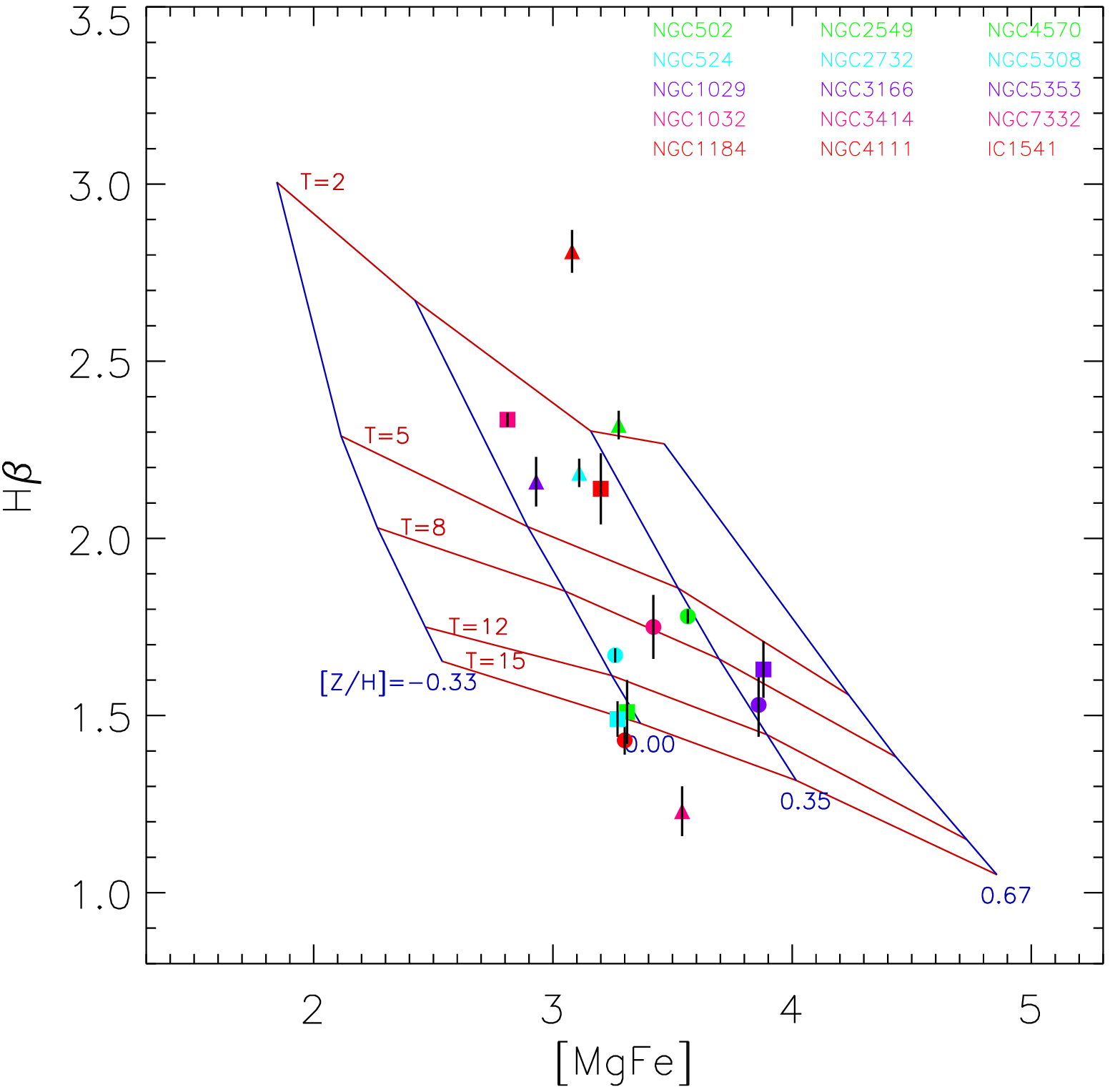} \\
 (a)&(b)\\
\end{tabular}
\caption{Diagnostic index-index diagrams for the bulges at $0.5r_e$ from the center.
Different galaxies are coded by different colours and different signs; in the
legend the first column contains galaxies coded by circles, the second column --
the galaxies coded by triangles, and the third column -- the galaxies coded by
squares.
{\bf (a)} --
The $\langle \mbox{Fe} \rangle$ vs Mgb diagram. The simple
stellar population models by \citet{thomod} for three different
magnesium-to-iron ratios (0.0, $+0.3$, and $+0.5$) and three different ages
(5, 8, and 12 Gyr) are plotted as reference. The small signs
along the model curves mark the metallicities of +0.35, 0.00,
--0.33, and --1.35, if one takes the signs from
right to left. {\bf (b)} -- The age-diagnostic diagram for the stellar
populations in the central parts of the galaxies under consideration;
the H$\beta$-index measurements are rectified
from the emission contamination where it is, as described in the text.
The stellar population models by \citet{thomod} for [Mg/Fe]$=+0.3$
and five different ages (2, 5, 8, 12 and 15 Gyr, from top to bottom curves)
are plotted as reference frame; the blue lines crossing the
model metallicity sequences mark the metallicities of +0.67, +0.35, 0.00,
--0.33 from right to left.}
\label{bulges}
\end{figure*}

To determine the stellar population properties, we compare our measured indices to
the models of Simple Stellar Populations (SSP) by \citet{thomod} for several
values of the magnesium-to-iron abundance ratio.
Figure~\ref{bulges} presents the index-index diagnostic diagrams which allow us to break
the age-metallicity degeneracy and to determine SSP-equivalent ages, metallicities, and
magnesium-to-iron abundance ratios for the stellar populations of the bulges which are
also listed in Table~4. As one can see, the bulges have mostly [Mg/Fe]$=+0.1 - +0.3\,$dex,
and a range of SSP-equivalent ages, from $<2$~Gyr to 15~Gyr. The total metallicities are solar
and higher.

\begin{figure*}
\hfil
\begin{tabular}{c c}
 \includegraphics[width=8cm]{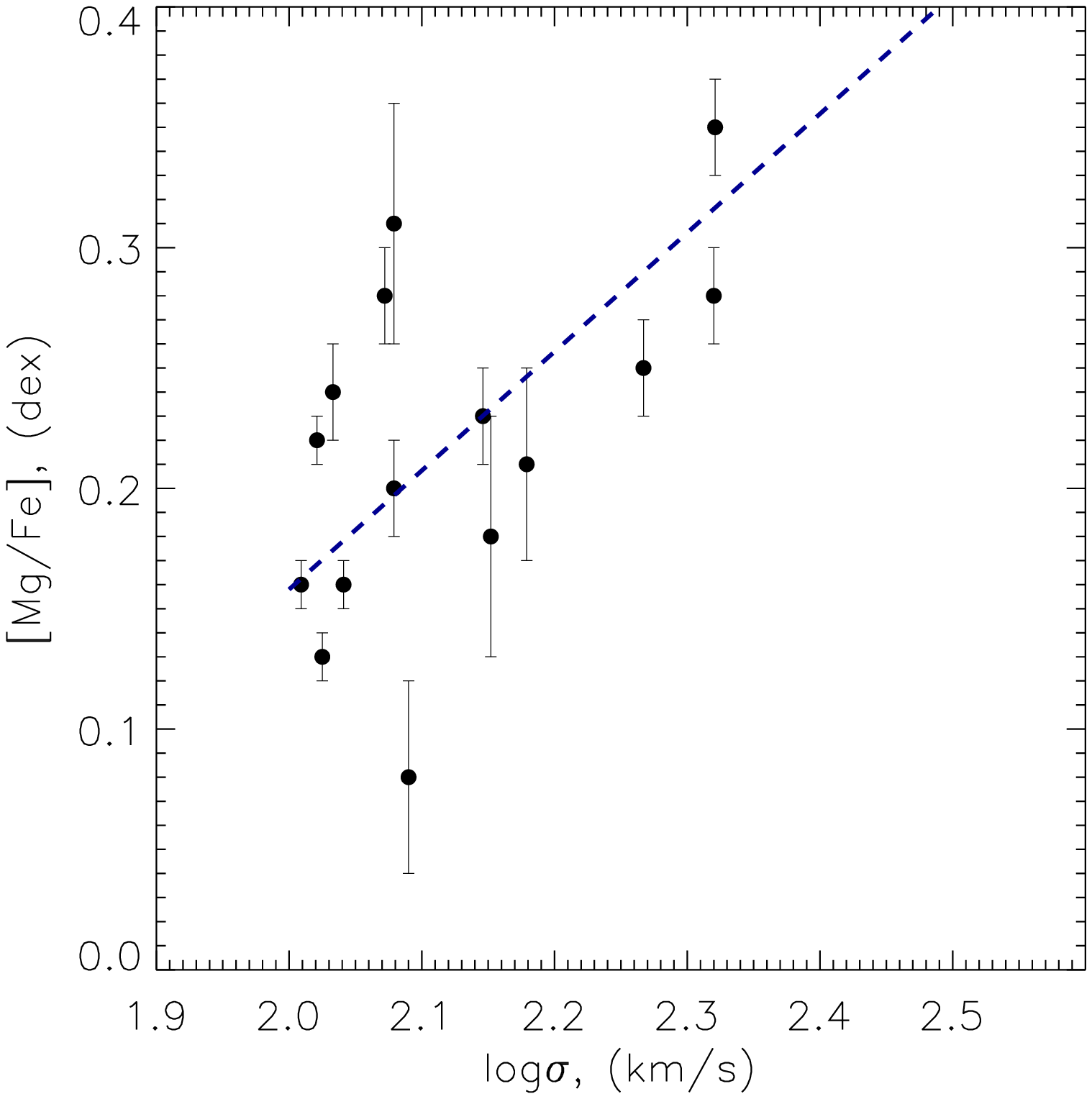} &
 \includegraphics[width=8cm]{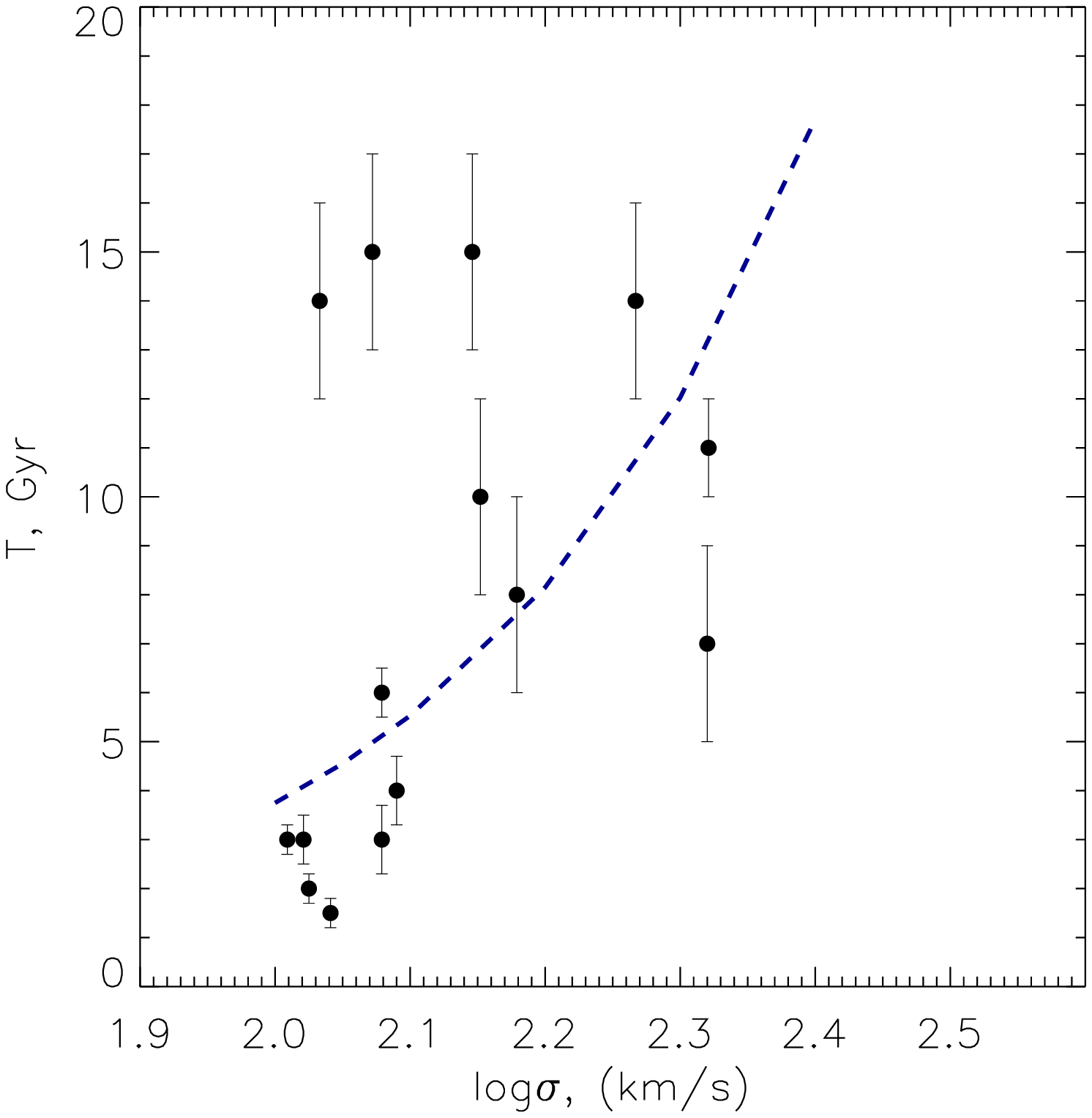} \\
(a) & (b) \\
\end{tabular}
\caption{The correlations found by us for the bulges of S0 galaxies, those
of the magnesium-to-iron ratio (a) and the age (b) versus the stellar velocity
dispersion  at the radius of $0.5r_e$ (the half effective radii of the
decomposed bulges); blue dashed lines present the regressions (the formulae
are in the text).}
\label{correl}
\end{figure*}

We have tried to find correlations between the parameters of the stellar populations of
the bulges and the stellar velocity dispersions at $R=0.5r_e$; the results are presented
in Table~6. In our small sample we have not found any correlation of the total metallicity
with the stellar velocity dispersion measured by us {\it at the same radius},
$0.5r_e$; marginal, just below $2\sigma$ correlations however exist between the 
age and $\sigma_*$ and between
[Mg/Fe] and  $\sigma_*$ (Fig.~\ref{correl}). Similar results were obtained by
Howell for the sample of nearby field ellipticals \citep{howell} and by us for
the sample of the bulges of 80 nearby S0s \citep{me2008}. For Fig.~\ref{correl}
we have calculated the regressions: [Mg/Fe]$=(0.49\pm 0.23)\log \sigma _* -(0.83\pm 0.50)$,
in full agreement with the recent results of \citet{annibalicor} for a sample of
early-type galaxies (ellipticals$+$ lenticulars) who give the slope of $0.42\pm 0.22$
and the zero point of $-0.74\pm 0.46$. However, our regression for the age,
$\log T (yr)=(1.69\pm 0.75)\log \sigma _* -(6.2\pm 1.6)$, is much steeper than
that by \citet{annibalicor}.

\subsection{The disks}

\begin{figure*}
\hfil
\begin{tabular}{c c}
 \includegraphics[width=8cm]{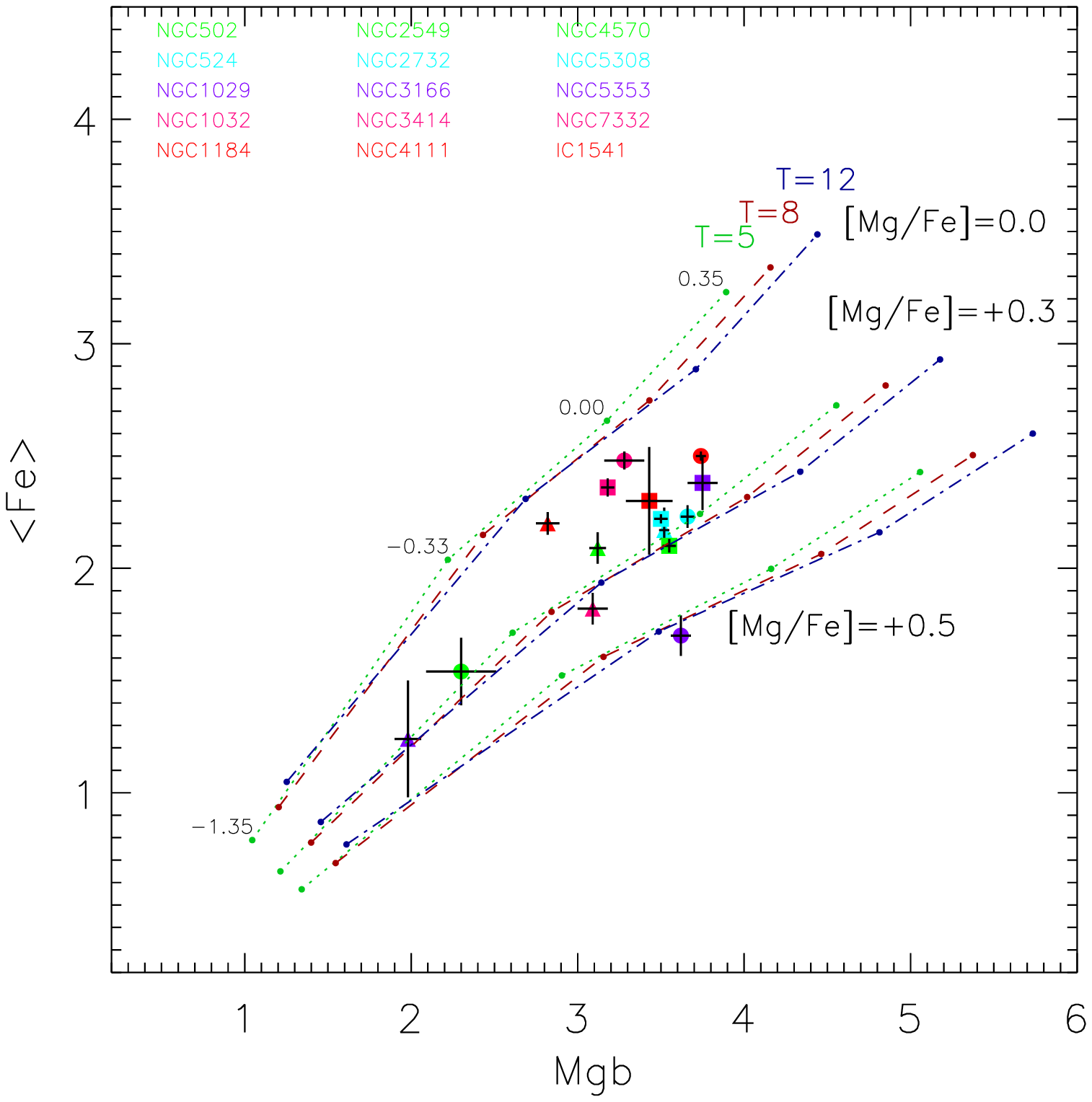} &
 \includegraphics[width=8cm]{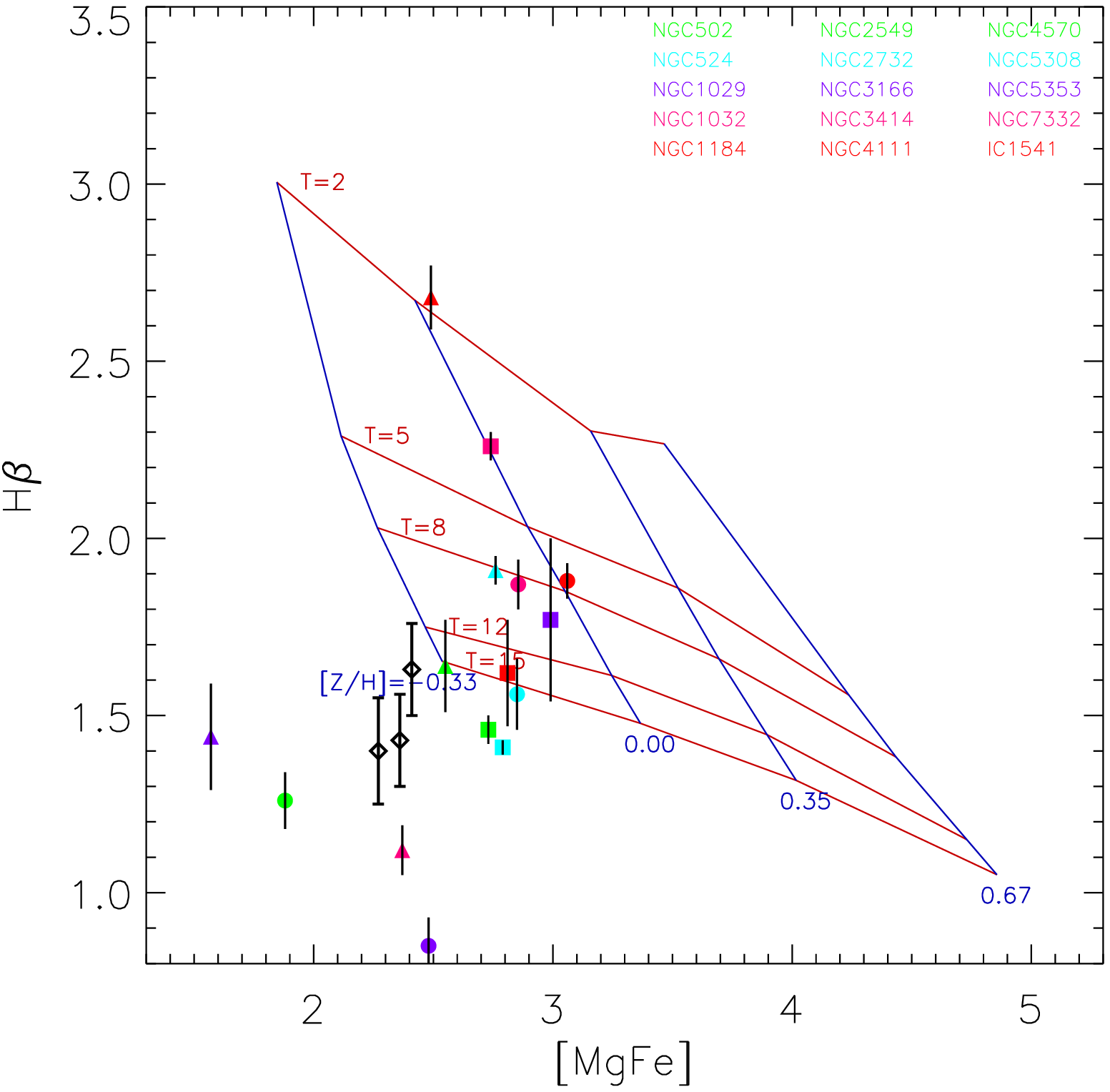} \\
 (a)&(b)\\
\end{tabular}
\caption{Diagnostic index-index diagrams for the large-scale stellar disks
averaged over their full extension.
The different galaxies are coded by different colours and different signs; in the
legend the first column contains galaxies coded by circles, the second column --
the galaxies coded by triangles, and the third column -- the galaxies coded by
squares.
{\bf (a)} --
The $\langle \mbox{Fe} \rangle$ vs Mgb diagram. The simple
stellar population models by \citet{thomod} for three different
magnesium-to-iron ratios (0.0, $+0.3$, and $+0.5$) and three different ages
(5, 8, and 12 Gyr) are plotted as reference. The small signs
along the model curves mark the metallicities of +0.35, 0.00,
--0.33, and --1.35, if one takes the signs from
right to left. {\bf (b)} -- The age-diagnostic diagram for the stellar
populations in the large-scale disks of the galaxies under consideration.
The stellar population models by \citet{thomod} for [Mg/Fe]$=+0.3$
and five different ages (2, 5, 8, 12 and 15 Gyr, from top to bottom curves)
are plotted as reference frame; the blue lines crossing the
model metallicity sequences mark the metallicities of +0.67, +0.35, 0.00,
--0.33 from right to left. Three globular clusters of our Galaxy, with intermediate
metallicities of [Fe/H]$=-0.4 - -0.7$, are also plotted by black diamonds for comparison;
their indices are taken from \citet{beasglob}.}
\label{disks}
\end{figure*}

We tried to probe the outer parts of the disks where the bulge contributions are
negligible and where there is no ionized gas so the emission line H$\beta$ does not
complicate the stellar age estimate. For NGC~502, we have summed two cross-sections
at {\it different} position angles, after assuring that because of the symmetrical
slit orientations with respect to the kinematical major axis they have identical
stellar LOS velocity profiles and that the Lick index radial distributions do not have systematic
shifts relative each other. The only complicated case was NGC~3414 where we were restricted
to the inner part of the antitruncated disk, $R<60\arcsec$, which is strongly polluted by the
Balmer emissions. We tried to correct the Lick index H$\beta$ using our procedure (as described above),
but perhaps some residual contamination has remained. Finally, we averaged the Lick index measurements
in the disks over rather extended radial intervals where the disks dominate photometrically
and where the index profiles look rather flat. These intervals are indicated in the Table~5,
as well as the mean Lick indices with their errors. As before, the errors are errors of the
means calculated from 3--7 individual points.

\begin{table*}
\scriptsize
\caption[ ] {Stellar population parameters of the disks}
\begin{flushleft}
\begin{tabular}{lrrrrrrrr}
\hline\noalign{\smallskip}
NGC/IC & Scale length $h\arcsec$ (Band), & Radius range, $\arcsec$ & H$\beta$ & Mgb & $\langle \mbox{Fe} \rangle$ &
$T$,~Gyr & $\mbox{[Z/H]}$ & $\mbox{[Mg/Fe]}$ \\
\hline
N502 & 10(V),9.5(r) & 20--30 & 1.26 & 2.30 & 1.54 & $>12$ & $<-0.5$ & $+0.20$ \\
 & & & $\pm 0.08$ & $\pm 0.21$ & $\pm 0.15$ & & & $\pm 0.08$ \\
N524 & 9(V) & 25--65 & 1.56 & 3.66 & 2.23 & $15\pm 3$ & $-0.2$ & $+0.30$ \\
 & & & $\pm 0.10$ & $\pm 0.04$ & $\pm 0.05$ & & & $\pm 0.02$ \\
N1029 & 9(K) & 20--35 & 0.85 & 3.62 & 1.70 & $>12$ & $<-0.4$ & $+0.46$ \\
 & & & $\pm 0.08$ & $\pm 0.06$ & $\pm 0.09$ & & & $\pm 0.02$ \\
N1032 & 29(r) & 40--60 & 1.87 & 3.28 & 2.48 & $9\pm 2$ & $-0.1$ & $+0.12$ \\
 & & & $\pm 0.07$ & $\pm 0.12$ & $\pm 0.04$ & & & $\pm 0.05$ \\
N1184 & 27(r) & 40--60 & 1.88 & 3.74 & 2.50 & $7 \pm 1$ & 0.0 & $+0.20$ \\
 & & & $ \pm 0.05$ & $\pm 0.03$ & $\pm 0.02$ & & &  $\pm 0.01$ \\
N2549 & 24(r) & 40--75 & 1.64 & 3.12 & 2.09 & $15\pm 4$ & $-0.3$ & $+0.26$ \\
  & & & $\pm 0.13$ & $\pm 0.05$ & $\pm 0.07$ & & & $\pm 0.05$ \\
N2732 & 10(R) & 25--45 & 1.91 & 3.52 & 2.17 & $8\pm 1$ & $-0.1$ & $+0.29$ \\
 & & & $\pm 0.04$ & $\pm 0.03$ & $\pm 0.10$ & & & $\pm 0.04$ \\
N3166 & 49(z),62(r) & 40--60$^{**}$ & 1.44 & 1.98 & 1.24 & $>12$ & $<-0.5$ & $+0.26$ \\
 & & & $\pm 0.15$ & $\pm 0.08$ & $\pm 0.26$ & & & $\pm 0.11$ \\
N3414 & 17(r) & 25--60 & 1.12 & 3.09 & 1.82 & $>12$ & $-0.4$ & $+0.33$ \\
 & & & $\pm 0.07$ & $\pm 0.09$ & $\pm 0.07$ & & & $\pm 0.05$ \\
N4111 & 28(r) & 40--70$^*$ & 2.68 & 2.82 & 2.20 &  $2\pm 0.3$ & $0.0$ & $+0.19$ \\
 & & & $\pm 0.09$ & $\pm 0.07$ & $\pm 0.05$ & & & $\pm 0.04$ \\
N4570 & 25(r) & 30--80 & 1.46 & 3.55 & 2.10 & $15\pm 1$ & $-0.2$ & $+0.35$ \\
  & & & $\pm 0.04$ & $\pm 0.04$ & $\pm 0.03$ & & &  $\pm 0.02$ \\
N5308 & 20(r) & 30--50 & 1.41 & 3.50 & 2.22 & $15\pm 1$ & $-0.2$ & $+0.29$ \\
 & & & $\pm 0.02$ & $\pm 0.04$ & $\pm 0.02$ &  & & $\pm 0.02$ \\
N5353 & 9(r) & 35--50$^*$ & 1.77 & 3.75 & 2.38 & $10\pm 5$ & $-0.1$ & $+0.36$ \\
 & & & $\pm 0.23$ & $\pm 0.09$ & $\pm 0.12$ & & & $\pm 0.07$ \\
N7332 & 22(r) & 30--70 & 2.26 & 3.18 & 2.36 & $3.5\pm 0.3$ & 0.0 & $+0.18$ \\
 & & & $\pm 0.04$ & $\pm 0.04$ & $\pm 0.04$ & & & $\pm 0.03$ \\
I1541 & 7(V) & 10--20 & 1.62 & 3.43 & 2.30 & $15\pm 5$ & $-0.2$ & $+0.24$ \\
 & & & $\pm 0.15$ & $\pm 0.14$ & $\pm 0.24$ & & & $\pm 0.14$ \\
\hline
\multicolumn{9}{l}{$^*$\rule{0pt}{11pt}\footnotesize
Only the southern part of the disk is measured.}\\
\multicolumn{9}{l}{$^{**}$\rule{0pt}{11pt}\footnotesize
Only the eastern part of the disk is measured.}\\
\end{tabular}
\end{flushleft}
\end{table*}

Figure~\ref{disks} presents the diagnostic index-index diagrams where our measurements
for the large-scale disks of the sample lenticular galaxies are compared to the SSP
models by \citet{thomod}. One can see that the stellar disks of our lenticulars are mostly
old, $T_{SSP}=8$~Gyr and older; only two galaxies, just those with young pseudobulges, 
NGC~4111 and NGC~7332, have also young large-scale stellar disks. The mean metallicities
of stellar populations in the disks are mostly subsolar -- only three youngest disks
reach solar metallicity; in NGC~502, NGC~1029, NGC~3166, and NGC~3414 the mean
disk stellar metallicities are very low, [Z/H]$=-0.4 - -0.7$, as we can see from comparison
of their location at the age-diagnostic diagram with some globular clusters of our Galaxy
(the SSP models for such metal-poor systems are perhaps not particularly good). And the most striking
feature of the large-scale disks in our lenticular galaxies is their high magnesium-to-iron
abundance ratio: in every galaxy but NGC~1032 the [Mg/Fe] of the disk is $+0.2 - +0.5$.
If we compare these stellar population characteristics to those of stars of
our Galaxy, they would resemble the thick disk of the MW \citep{fuhrmann08},
with its age above 10~Gyr, subsolar stellar metallicities and
$\alpha$-elements overabundance.

Concerning the usually probed correlations, there is certainly no correlation between 
the metallicity or [Mg/Fe] ratio and
the mass characteristics of the disks, $V_{rot}^2 + \sigma _* ^2$ (Table~6).
For these particular correlations, we have analysed only edge-on galaxies, so here the
stellar velocity dispersions measured along the line of sight are close to the
tangential components of the stellar chaotic motions.

\begin{figure*}
\hfil
\begin{tabular}{c c}
 \includegraphics[width=8cm]{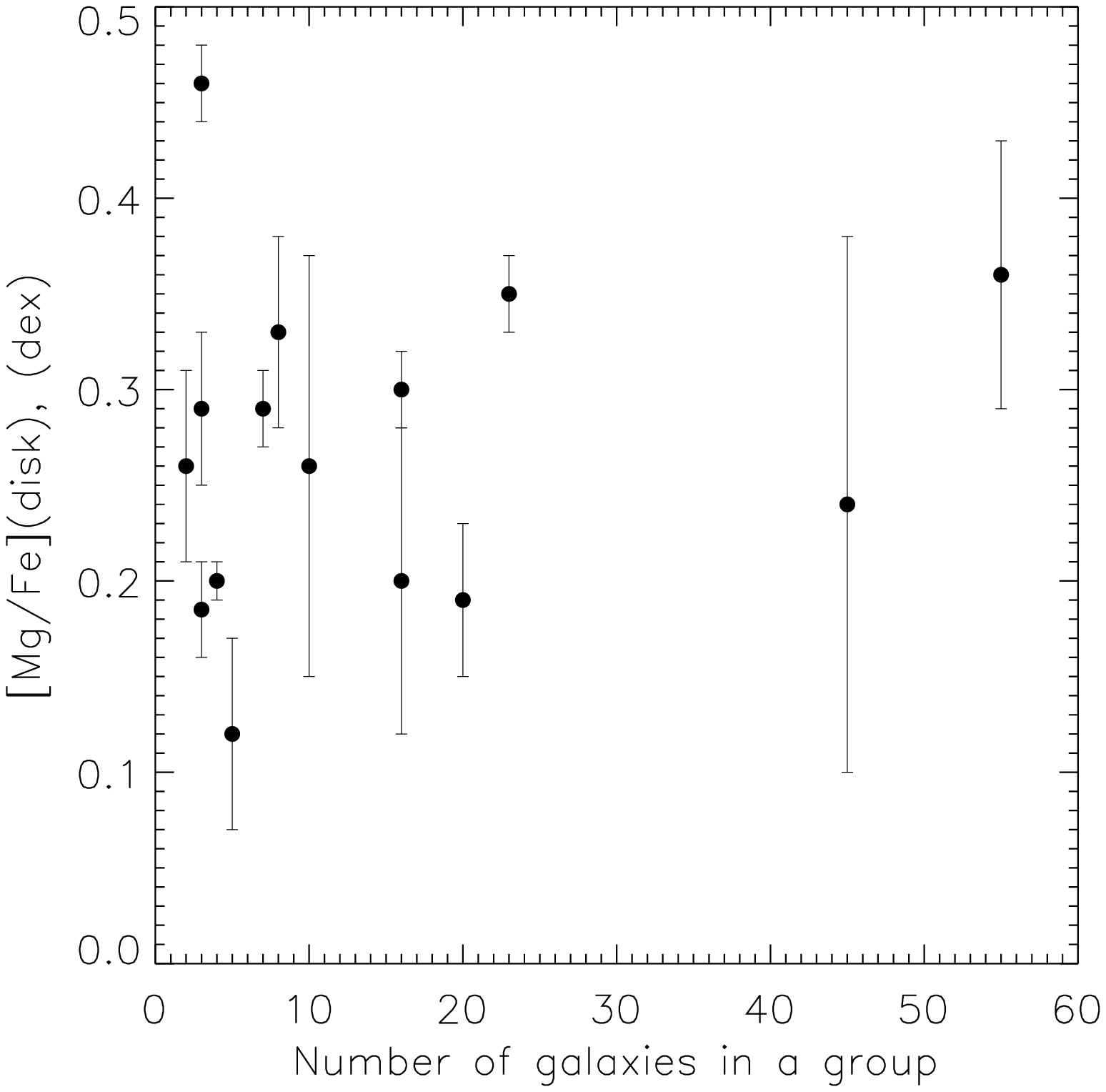} &
 \includegraphics[width=8cm]{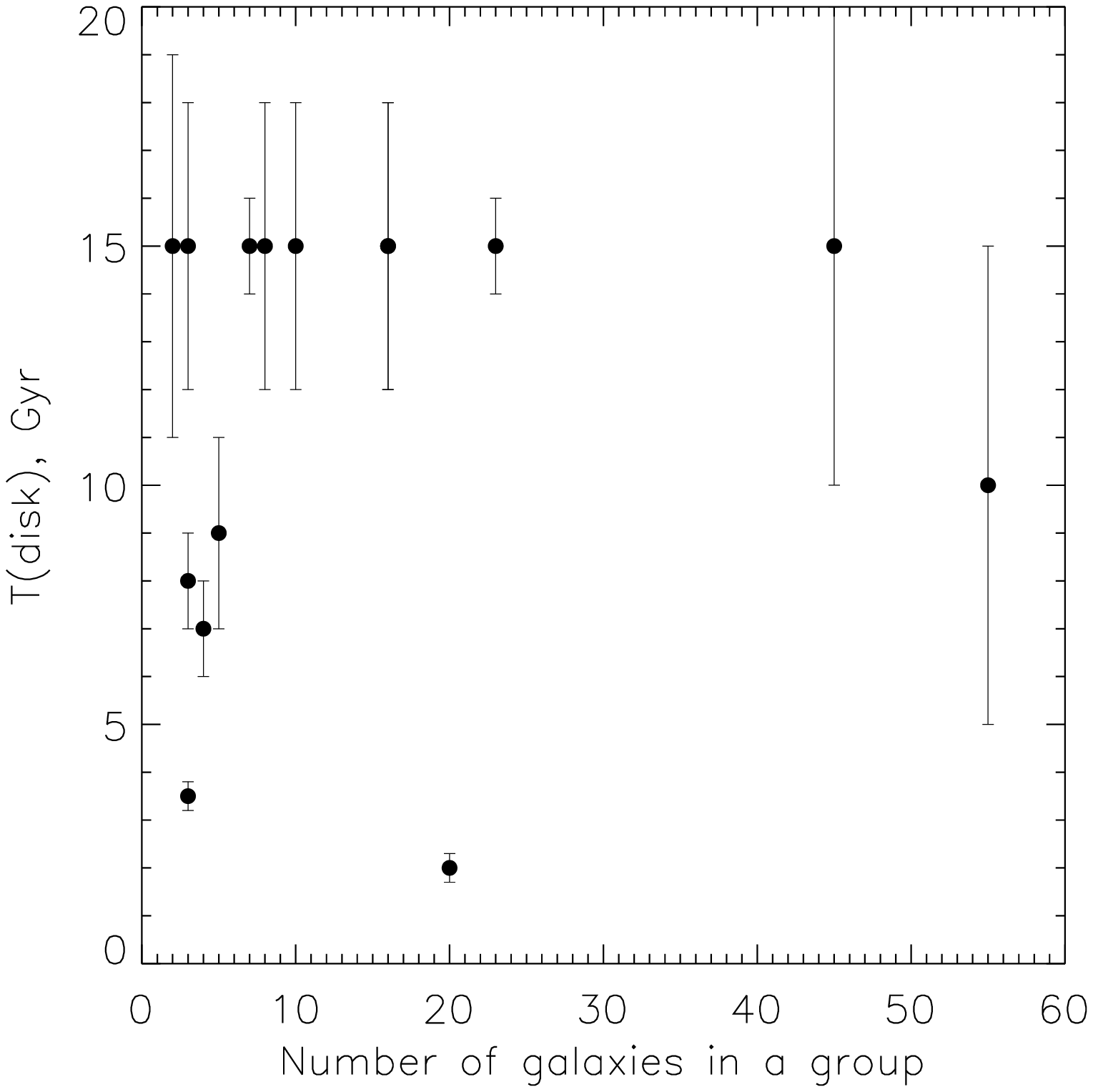} \\
(a) & (b) \\
\end{tabular}
\caption{The correlations found by us for the disk stellar population properties 
of the sample S0 galaxies and their environment density (the number of
galaxies in the groups to which the sample S0s belong).}
\label{correlenv}
\end{figure*}

Another set of possible correlations may exist between the properties of the disk stellar
populations and the environment density. To quantify the environment density, we
took the numbers of galaxies in the groups to which the sample galaxies
belong. To make the estimates as homogeneous as possible, we used the recent
group catalogue by \citet{makkar} for most galaxies; only NGC~1029 which is
absent in the catalogue by \citet{makkar} is attributed according to the catalogue
of isolated triplets by \citet{triplets}, and for IC~1541, which is too far from us
to be involved into the catalogue by \citet{makkar}, we have used the estimates from
the work by \citet{mahdavigeller}. Figure~\ref{correlenv} presents the plots  of
the disk ages and magnesium-to-iron ratios versus the environment density.  
For the age, the correlation is surely absent. We can only conclude from the right plot 
of Fig.~\ref{correlenv} that in sparse environments the disk age estimates cover the full
range of possible values; in other words, the S0 galaxies in sparse environments
sometimes rejuvenated their disks after $z<1$.

\begin{table*}
\caption[ ]{Correlations between stellar population properties
and mass characteristics of galaxy components}
\begin{flushleft}
\begin{tabular}{|c|c|c|}
\hline\noalign{\smallskip}
 & Spearman's correlation coefficient & Probability of no correlation \\
\hline
\multicolumn{3}{|l|}{Bulges, at $r=0.5r_{eff}$, 15 galaxies}\\
$T$ {\bf vs} $\log \sigma _*$ &    0.500 & 0.058 \\
 $[\mbox{Z/H}]$ {\bf vs} $\log \sigma _*$ & --0.069 & 0.808 \\
 $[\mbox{Mg/Fe}]$ {\bf vs} $\log \sigma _*$ & 0.462 & 0.083 \\
\hline
\multicolumn{3}{|l|}{Disks, only edge-on, 11 galaxies}\\
$T$ {\bf vs} $\log (v_{rot}^2 + \sigma _* ^2)$ & 0.380 & 0.248 \\
 $[\mbox{Z/H}]$ {\bf vs} $\log (v_{rot}^2 + \sigma _* ^2)$ & -0.234 & 0.489 \\
 $[\mbox{Mg/Fe}]$ {\bf vs} $\log (v_{rot}^2 + \sigma _* ^2)$ & 0.251 & 0.456 \\
\hline
\end{tabular}
\end{flushleft}
\end{table*}

\begin{figure*}
\hfil
\begin{tabular}{c c}
 \includegraphics[width=8cm]{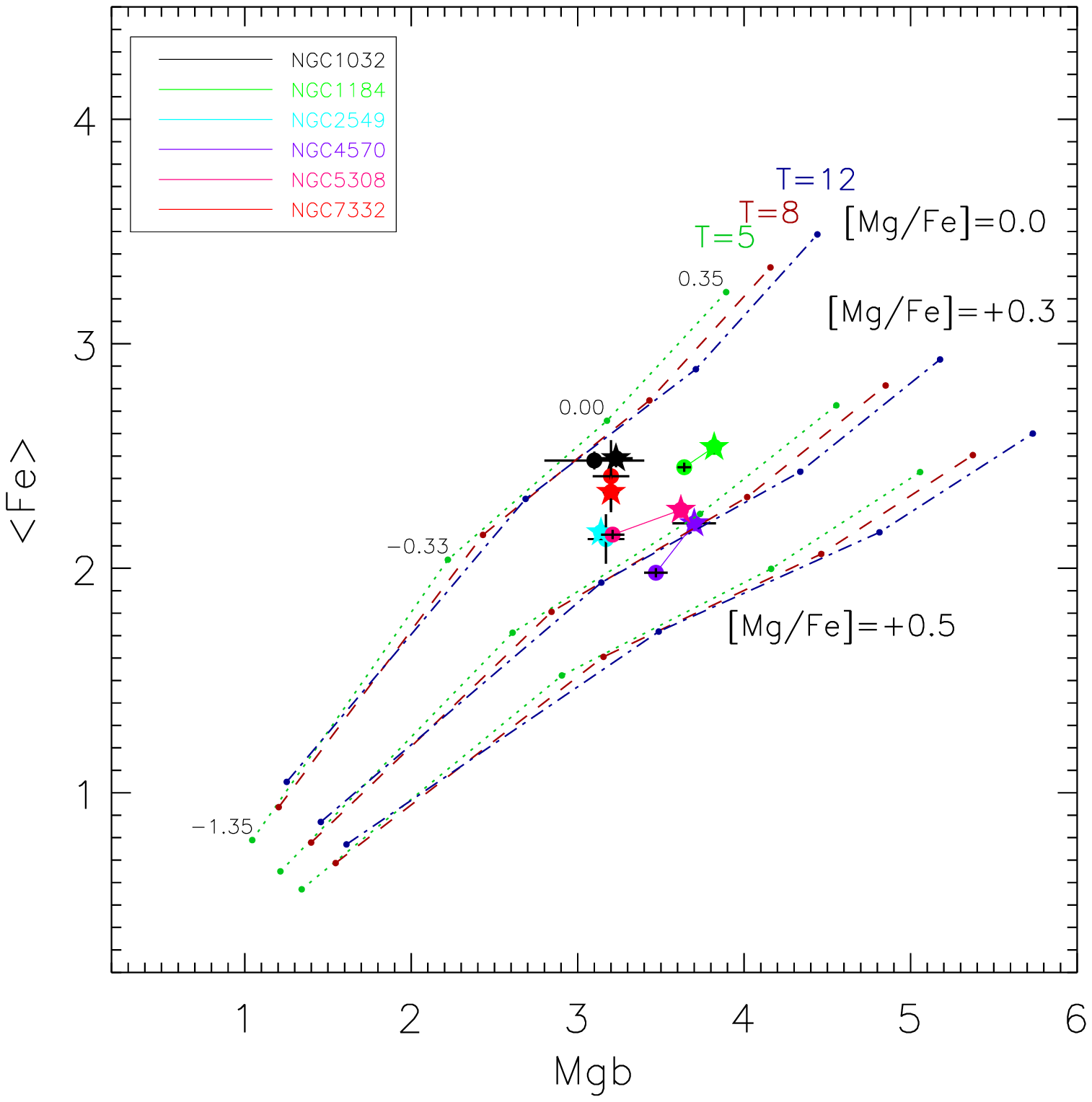} &
 \includegraphics[width=8cm]{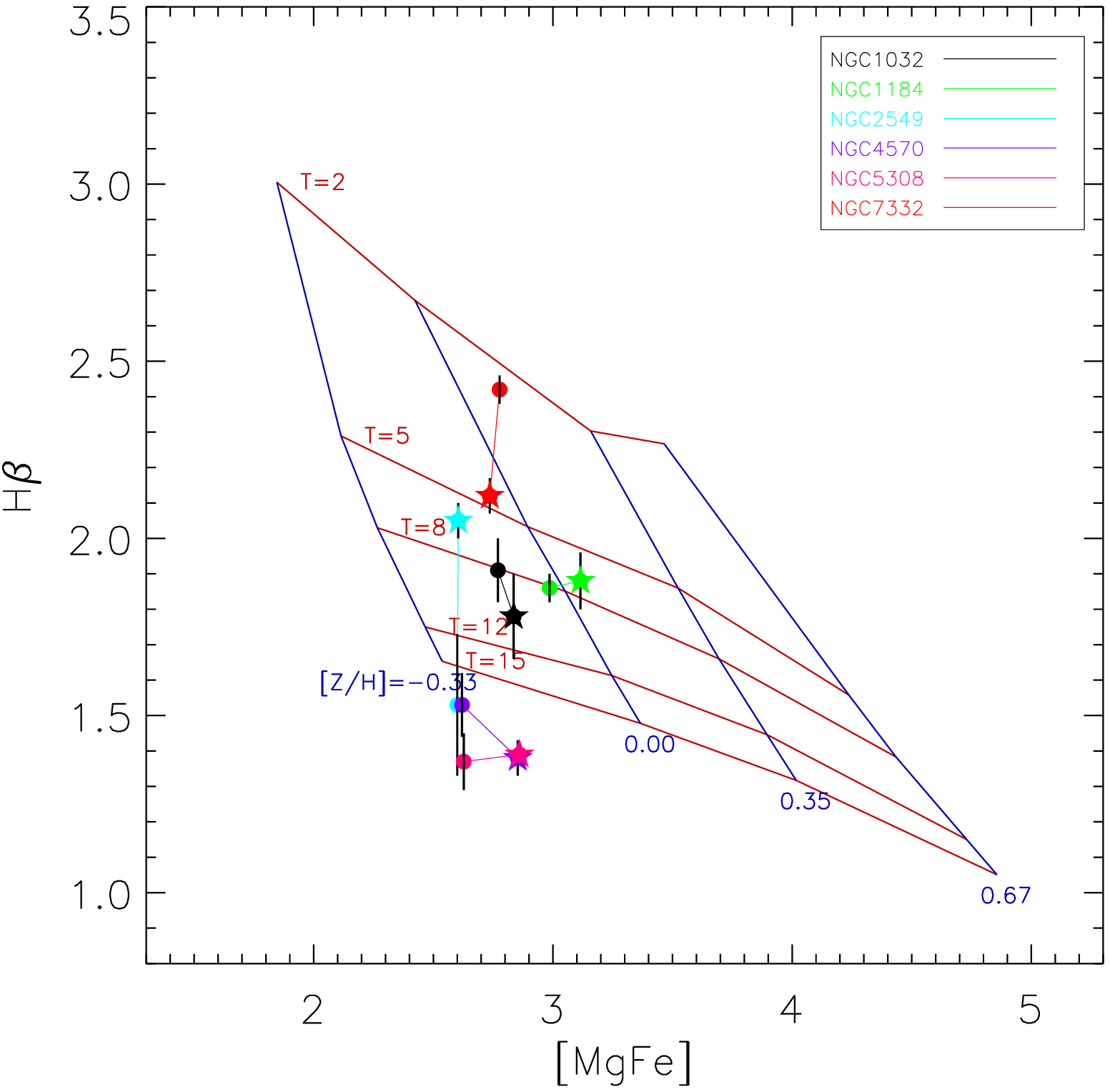} \\
 (a)&(b)\\
\end{tabular}
\caption{Diagnostic index-index diagrams for disks as a function of the distance from the galaxy centers
{\bf (a)} --
The $\langle \mbox{Fe} \rangle$ vs Mgb diagram. The large signs connected by lines
are our galaxies' disks with the innermost measurements marked by stars and
the outermost measurements marked by filled circles. The simple
stellar population models by \citet{thomod} for three different
magnesium-to-iron ratios (0.0, $+0.3$, and $+0.5$) and three different ages
(5, 8, and 12 Gyr) are plotted as reference. The small signs
along the model curves mark the metallicities of +0.35, 0.00,
--0.33, and --1.35, if one takes the signs from
right to left. {\bf (b)} -- The age-diagnostic diagram for the stellar
populations in the disks of the galaxies under consideration.
The large signs connected by lines
are our galaxies' disks with the innermost measurements marked by stars and
the outermost measurements marked by filled circles.
The stellar population models by \citet{thomod} for [Mg/Fe]$=+0.3$
and five different ages (2, 5, 8, 12 and 15 Gyr, from top to bottom curves)
are plotted as reference frame; the blue lines crossing the
model metallicity sequences mark the metallicities of +0.67, +0.35, 0.00,
--0.33 from right to left.}
\label{grads}
\end{figure*}

Several galaxies have rather extended Lick index profiles, and we are able to estimate
stellar population parameters gradients {\it within the disks} (Fig.~\ref{grads}).
Again we see a difference between the galaxies in dense environments having old disks
and the galaxies in sparse environments having intermediate-age disks.
In NGC~5308 and NGC~4570 the disks are homogeneously old at all radii, and the metallicity
gradient between $R=2.5$ kpc
and $R\approx 6$ kpc (if we assume that NGC~4570 is at the distance of the Virgo cluster)
does not exceed  --0.04 dex per kpc that is quite typical also for
early-type spiral galaxies \citep{gradmet}.
The isolated galaxy NGC~1184 shows negligible gradients both in age and
abundances. In NGC~7332, the member of a non-interacting triplet,
and in NGC~1032, isolated, the mean stellar age
{\it falls} along the radius, from 5 to 2.5~Gyr in the former and from 10 to 8~Gyr in the latter.
Consequently, the metallicity gradient is positive in NGC~7332 and zero in NGC~1032: the
prolonged star formation in the outer disks have increased there the mean stellar metallicity.
NGC~2549, the loose-group central galaxy, demonstrates quite a different behaviour:
its SSP-equivalent stellar age rises significatly between $R=40\arcsec$ ($\sim 3$ kpc) and
$R=70\arcsec$ ($\sim 5$ kpc), from $6.5\pm 0.5$ Gyr to $\ge 12$ Gyr, while the metallicity
decreases moderately, by 0.15 dex, so the metallicity gradient can be estimated
as --0.08 dex per kpc. Here we must keep in mind that NGC~2549 has a wide surface-brightness
excess -- probably, a starforming-ring relic -- at the radius of $R=25\arcsec -40\arcsec$
\citep{ss}.

In NGC~524 which has an antitruncated large-scale stellar disk \citep{n524gr} with
the high surface brightness inner part, we are able to estimate separately the parameters
of the stellar populations in the inner, $r=10^{\prime \prime} - 25^{\prime \prime}$,
and in the outer, $r=25^{\prime \prime} - 65^{\prime \prime}$, stellar disks. We
found H$\beta =1.62$ and [MgFe]$=2.85$ for the inner disk, so the parameters of the
stellar populations are $T=13\pm 2$~Gyr and [Z/H]$=-0.1$. We can state that the inner
disk looks slightly younger and slightly more metal-rich than the outer one, though
the difference is within the accuracy of our measurements.

\section{Discussion}

The old ages and strong magnesium overabundances of the large-scale stellar disks measured by us
in this work for the sample of S0s contradict the commonly accepted paradigm described in the
Introduction -- that S0 galaxies have born `en mass' from spiral
progenitors by quenching star formation in their disks when having fallen
into dense environments around $z=0.4$ (4~Gyr ago). 
In meantime, the SSP-equivalent age of 8~Gyr which is found for the disks
of the field galaxies NGC~1032, NGC~1184, NGC~2732 may reflect the constant-rate star
formation with quenching abruptly 5~Gyr ago; the SSP-equivalent age of 12~Gyr
corresponds to the constant-rate star formation with quenching 10~Gyr ago (Smith et al.
2009, Allanson et al. 2009), or at $z\sim 2$. If the star formation before quenching
was not constant but e-folding, the quenching must have happenned even earlier given
the SSP-equivalent ages for the S0 disks found by us \citep{smith2009,allan2009}.
IC~1541, NGC~502, 524, 3414, 5308 (rich group members) and NGC~4570 (Virgo cluster member) have
their SSP-equivalent stellar ages of the large-scale disks older than 12~Gyr, so they quenched
their star formation at $z>2$. It seems that the sudden appearance of (red, bulge-dominated)
S0 galaxies in clusters and groups at $z=0.4$, or 4~Gyr ago, is not related to their
morphological shaping. Within these dense environments the lenticulars,
shaped earlier and accreted during the massive halo assembly at $z<1$, might experience some
brief rejuvenation of their inner parts (bulges) forcing them to look  bluer at $z=0.4-0.7$
than later (and earlier). This idea was proposed by \citet{burstein} who did not find
the required large difference between typical luminosities of S0s and spirals
that was expected in the scenario of S0s fading from spirals.
At $z>0.7$ the same galaxies/bulges look exactly as
red as at $z=0$ \citep{kooclust,koofield}. In the frame of this hypothesis, the numerous
results on early-type galaxies number increase in clusters and groups when moving from
$z=0.8$ toward $z=0$ (e.g. \citet{simard09}) can be easily understood. Early morphological
types are often classified quantitatively  by the high $B/T$ ratio or by the high Sersic
index when applied to a galaxy as a whole. So the natural secular bulge build-up of disk galaxies
within dense environments under gravitational and hydrodynamical influences should provide
just a visibility of the early-type galaxies number increase with time. In reality, the main
morphological attribute of S0 galaxies -- large-scale old stellar disks -- is already in place
at $z=0.8-1$, though the whole population of S0s at $z=1$ may possess on average smaller bulges
than the present S0 population. Curious dependencies found by \citet{dressler}, namely,
the bulge {\it size} increasing with the environment density demonstrated both by S0s and spirals
(and the steeper increase is demonstrated just by spirals!), are quite in line with this idea.

\begin{figure}
\centering
\includegraphics[width=0.33\textwidth]{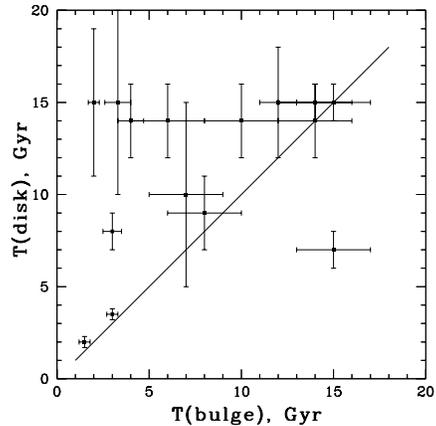}
\caption{The comparison of the obtained SSP-equivalent ages of the disks and
bulges in our sample S0s. The equality straight line is plotted for the reference.}
\label{agecomp}
\end{figure}

Figure~\ref{agecomp} presents the comparison of the SSP-equivalent ages between the disks and
the bulges in the sample S0s. We see that while the disks are mostly old, the bulges occupy
uniformly the whole range of ages between 2 and 15 Gyr; and the disks are almost always
older than the bulges. The only exception is NGC~1184, rather isolated galaxy, which has
the old bulge and the disk of intermediate age. We can also mention an analogous case of
NGC~3115 -- the quite isolated S0 galaxy which has the 12~Gyr old bulge and the 6~Gyr old disk
\citep{n3115}. The relation between the bulge and disk ages similar to our
Fig.~\ref{agecomp} has been obtained by \citet{prchamb} with their sample
of 59 S0s. They constructed a distribution of $\Delta T(\mbox{disk-bulge})$ which
looked like a Gaussian peaked at zero with a long tail to positive values. Also,
24\%\ of their sample galaxies have revealed the outer disks older than 10~Gyr. In principle,
almost all the mechanisms proposed up to date to transform a spiral galaxy into S0 are able
to produce star formation bursts just in the centers of the galaxies.
Hydrodynamical mechanisms acting through stripping cold gas from the outer disks by hot-gas
ram pressure leave intact the inner gas of the galaxies infalling into a cluster or a massive
group and even compress it provoking star formation bursts \citep{quilis,kronmodel}.
Recent simulations by \citet{bekkicouch} of a spiral transforming into S0 by tidal
distortions inside a galaxy group show again multiple starbursts in the bulge area during
the transformation. But to upbuild a disk, or to burn secondary star formation over an
extended disk-dominated area, a galaxy needs obviously to settle within sparse
environments where smooth cold-gas accretion is possible. This idea has an observational
support: for example, blue-cloud E/S0s which demonstrate star formation over the whole
galaxy body at the present epoch, are located in sparse environments \citep{kannap}.

\begin{figure*}
\hfil
\begin{tabular}{c c}
 \includegraphics[width=8cm]{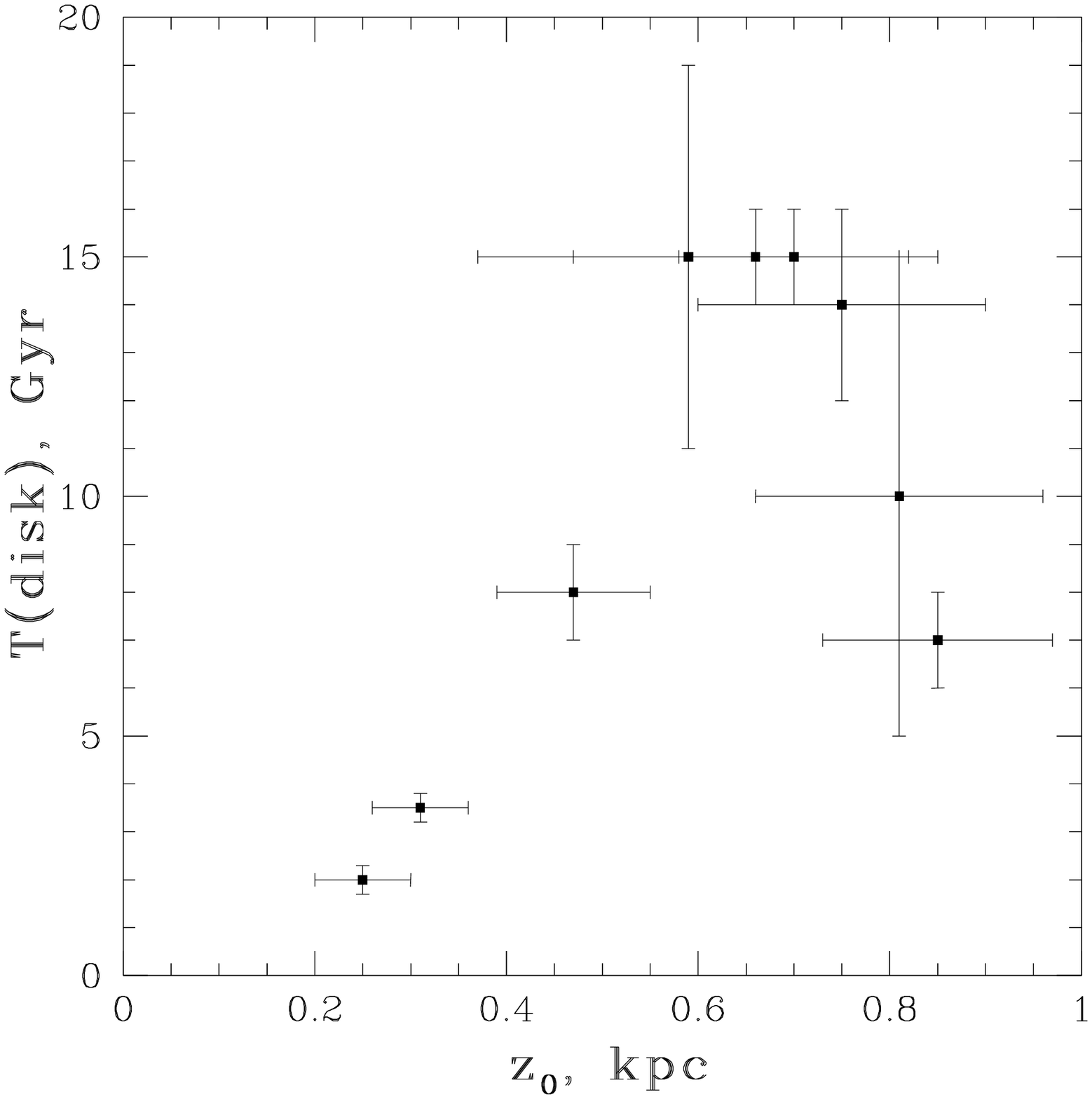} &
 \includegraphics[width=8cm]{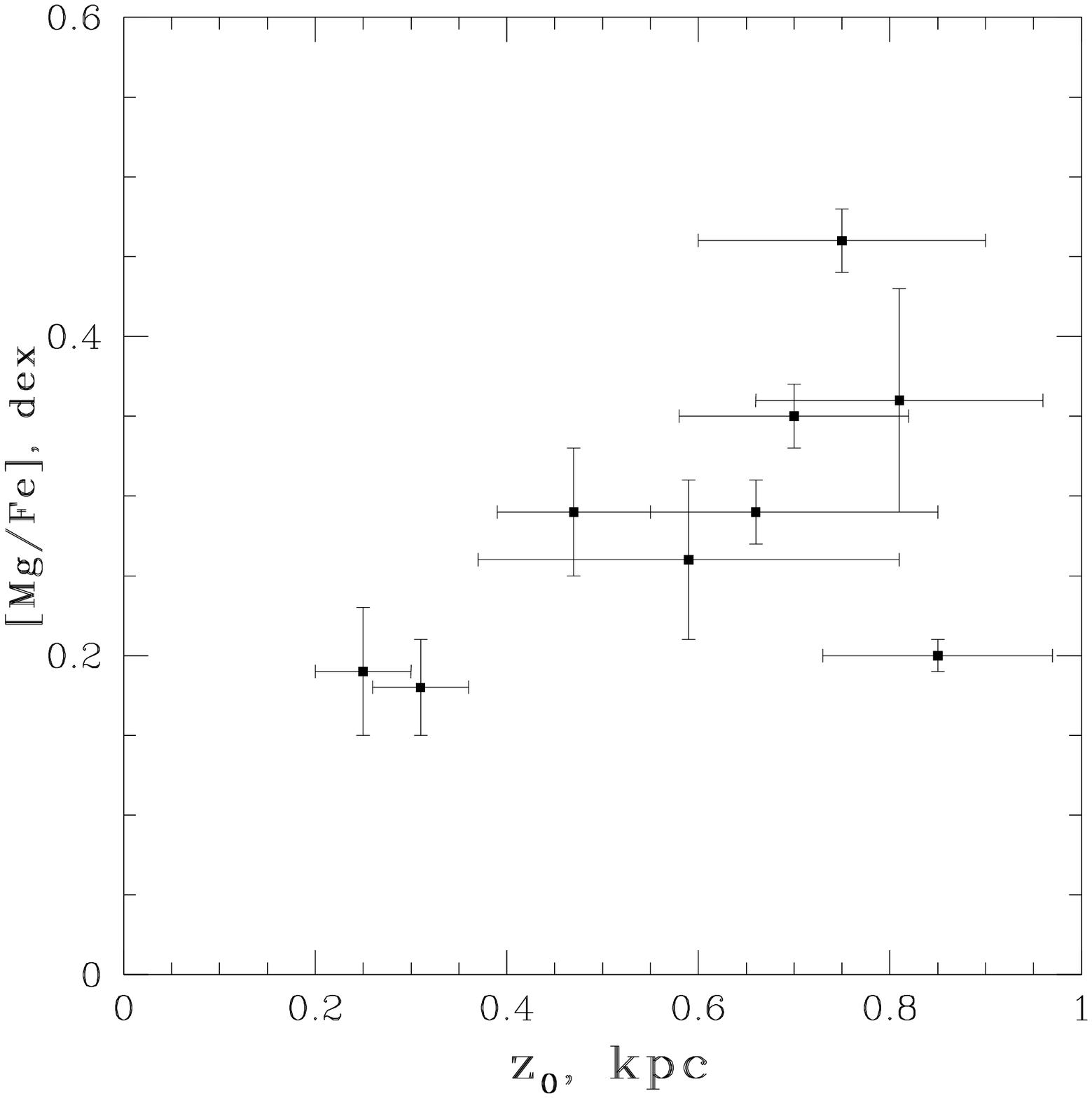} \\
 (a)&(b)\\
\end{tabular}
\caption{Correlations between the scale heights of the stellar disks derived
from the surface photometry of the edge-on S0 galaxies and their stellar age
({\bf a}) and magnesium-to-iron ratio ({\bf b}).
}
\label{vsheight}
\end{figure*}

As we have noted above, the stellar population chemistry
in the disks of lenticular galaxies of our sample appear to lack any correlation
with the mass (characterized by $\sigma ^2 + v_{rot}^2$) or with the luminosity of the disks.
However, some hints to correlations have been found, and these are correlations with the thickness 
of the disks. In Fig.~\ref{vsheight} we confront the ages and the magnesium-to-iron ratios
of the stellar populations in the disks of 9 edge-on S0s versus the scale heights
of the disks found by \citet{edgeonphot} through the surface photometry analysis.
The correlation between [Mg/Fe] and $z_0$ is suggestive, with the Spearman
correlation coefficient of 0.6 (less than 10\%--probability of no dependence).
The correlation between the age and the disk thickness is formally insignificant, 
with the Spearman correlation coefficient of about 0.35,
due to the thick disk of NGC~1184 which has an intermediate age; but neverthless
all the {\it old}  disks of our sample galaxies are {\it thick} disks,
while the only two young disks, those of NGC~4111 and NGC~7332, are
certainly {\it thin} disks. The disk of NGC~2732, with its SSP-equivalent age of 8~Gyr
and the scale height of 0.5 kpc, is halfway between thin and thick disks.
The range of the scale heights is from 0.3 kpc (in the disks with the ages of 2--3~Gyr
and [Mg/Fe]$\le +0.2$) to 0.6--0.9 kpc ( mostly in the disks with the ages $\ge 10$~Gyr
and [Mg/Fe]$\ge +0.3$). Let us compare it to our Galaxy disks' scale heights: 0.3 kpc for
the thin stellar disk and about 1 kpc for the thick stellar disk \citep{galdisksz}.

Many arguments evidence for S0s and spiral galaxies being relatives -- the most 
recent arguments in favour of the parallel morphological sequences of Ss and S0s can 
be found, e.g., in \citet{kormbend,atlas3d_7}. But after taking into account our present
results, it looks like S0s are progenitors of spirals, opposite to what was thought before.
Indeed, if we compare stars of the thick disk of our own Galaxy with the thick
stellar disks of S0s studied here we will see full resemblance: the ages $>10$~Gyr,
[Mg/Fe]$>+0.2$, the total metallicity, which is closer to [Mg/H] due to Mg coupling
with oxygen, both being $\alpha$-elements, than to [Fe/H], [Z/H] is between $0.0$ and $-0.7$
\citep{bernfuhr,schuster2006}. In other spiral galaxies thick stellar disks
are also much older than the embedded thin disks \citep{yd08}.
So if now one provides fresh cold gas accretion into the disks of our S0s, after several
Gyrs of star formation we would get typical spiral galaxies, with the thick old stellar disks
and thin younger stellar disks. The idea by Fuhrmann \citep{fuhrmann11} that primary
large-scale components of all galaxies must be thick stellar disks plays here nicely.

Indeed, observations reveal that star formation in disk galaxies at $z\sim 2$
proceeds in clumps with the sizes of about 1-1.5 kpc embedded into large-scale disks
which are gravitationally bound; and the scale heights of these disks correspond to
the clump sizes so the gaseous disks at $z\sim 2$ are thick \citep{bournaud08}. Hence, the
stellar disks forming from this gas in these galaxies should be also thick.
The SPH simulations of the secular evolution for such a configuration promise very effective
(and so brief) star formation \citep{bournaud07}; observations of starforming galaxies 
at $z>2$ confirm the short timescales of their star formation \citep{genzel};
so after ceasing star formation the emerged passively evolving stellar structures
would possess magnesium-overabundant stellar population. 
The first observations of the massive, disk-dominated, passive (no star formation) 
galaxy population at $z>2$ have already appeared in the literature \citep{brucetal}.

Initial simulations of the evolution of massive clumpy turbulent star-forming disks
implied strong radial inflow of the clumps into the galactic centers, and so it seemed
to be a way to form bulges of the future early-type disk galaxies: the resulting model bulges
looked similar to the `classical' bulges, with large Sersic indices and slow rotation \citep{elm08}. 
However later inclusion of the star formation feedback into the simulations has resulted
in stopping strong gas radial inflows due to shorter lifetimes of the clumps; and instead
of the bulges, thick stellar disks emerge now from these simulations \citep{genel,hopkins}.
\citet{bournaud09} conclude directly that intense star formation in high-redshift
clumpy disks may produce the present thick disks of spiral galaxies. 

So now we are proposing the following new scenario of disk galaxy evolution.
All disk galaxies were S0s immediately after their birth at $z>2$. Later, at $z<1$, some of them 
were provided with cold gas accretion sources to form thin stellar disks -- those might become
spirals, -- and some of them failed to find such sources -- these remained lenticulars.
Inside large cluster-size and group-size dark haloes, there are little chances to find
external sources of cold gas accretion, due to surrounding hot intergalactic gas, -- 
so in nearby clusters the dominant disk-galaxy population is lenticulars.
Or perhaps, the tidal effects -- harrassment resulting in starvation -- are more
effective in stripping the outer cold gas reservoirs of the disk galaxies
preventing late building of thin disks; this hypothesis is more in line with
the smooth dependence of the S0 (and S) fraction on the local environment density
over a full range of the latter parameter, from field to clusters \citep{pg84,atlas3d_7}. 
Our scenario implies that the disks of spiral galaxies must be on average more
massive (and so more luminous, say, in the $K$-band) than the disks of S0s; indeed, we see
such difference in Fig.~11 of the paper by \citet{lauri10} where, at fixed
bulge luminosity, the disks of spirals are more luminous than the disks of S0s.
Moreover, the bulges of early-type spirals are also on average slightly more
luminous (in the $K$-band) than the bulges of S0s \citep{grahamwor}; so we think that
the bulge growth during thin disk formation is probably inavoidable. In general, our 
scenario explains long-standing problem with the S0s being fainter than E and Sa galaxies
between which they are positioned at the Hubble's `tuning fork' \citep{vdb09}.
Also, it explains why the Tully-Fisher relation of S0 galaxies
goes in parallel to the Tully-Fisher relation for spirals, but with the 0.5 mag shift in 
the $K_S$-band toward fainter luminosities at the fixed rotation velocity \citep{williams}:
this shift is too small for the star formation truncation 4--5~Gyr ago but, as the authors
conclude, `could therefore be explained by a systematic difference between the total mass 
distributions of S0s and spirals, in the sense that S0s need to be smaller ... than spirals'.

The open question remains what can be these sources of cold-gas prolonged accretion  
-- they may be cosmologically motivated filaments \citep{dekel} or rich systems
of irregular-type dwarf satellites which have been merging with the host galaxy one after
another. To support the latter possibility, we would like to bring forward
the recent curious finding by \citet{mig}: by considering a sample of isolated
host galaxies, they have found that the pairs `S0$+$satellite' have on average twice higher
velocity dispersion (LOS velocity differences) than the pairs `S$+$satellite'.
A close inspection of their figures reveals an absence of the `S0$+$satellite' pairs with
the velocity differences less than 50 km/s. It might mean that the satellites of the present
S0s have much less chances to merge with their host galaxies than the satellites
of spirals. Perhaps, among a primordial population of isolated S0s covering a full 
range of satellite velocity dispersions the host galaxies with the lower satellite 
velocity dispersions have become spirals, and only the host galaxies with the high
satellite velocity dispersions preventing their numerous minor mergers remain still S0s.

\section{Conclusions}

We have studied the stellar population properties along the radius up to several
scale lengths of the disks in 15 S0 galaxies spread over a range of luminosities
(though all are more luminous than $M_B=-18$ and $M_K=-22$) and settling in different environments.

For the large-scale stellar disks of the galaxies, we have found metallicities from 
the solar one down to [Z/H]$=-0.4 - -0.7$, elevated magnesium-to-iron ratios,
[Mg/Fe]$\ge +0.2$, and mostly old ages. Nine of 15 galaxies have large-scale disks
older than 10~Gyr, which includes all the galaxies from the sample which reside
in dense environments.
The isolated and some loose-group galaxies have intermediate-age(7-8~Gyr) stellar disks,
and only two galaxies, NGC~4111 and NGC~7332, demonstrate the SSP-equivalent ages
of their disks of 2--3~Gyr. Just these two young disks have appeared to be thin,
and the other, old, disks have scale heights typical for thick stellar disks.

We conclude that S0 galaxies are the primordial type of disk galaxies 
largely shaped by $z\sim 1.5-2$, especially those found in clusters today.
Some rejuvenation of the disks, but mostly of the bulges,
is possible later for some, but not all, lenticular galaxies. The bulges are almost
always coeval or are younger than the disks. Spiral galaxies may form from
lenticulars at $z\le 1$ by accreting external cold gas into their disks; in
dense environments cold gas deficiency exists so the most primordial lenticulars
in dense environments remain S0s up to now being the dominant galaxy population
in the galaxy clusters at $z=0$.

Despite apparent homogeneity of the S0 morphological type (`smooth red disk$+$bulge systems...'), 
the glance `in depth' reveals very large scatter of main structural and evolutionary properties even 
among lenticular galaxies of the same luminosity: S0s may be disk- or bulge-dominated galaxies
\citep{vdb76,kormbend}, with exponential or de Vaucouleurs'-like surface 
brightness profiles of the bulges, and all the intermediate values of the Sersic parameter $n$
can be met too \citep{lauri10}; and the stellar population of the bulges spans the full 
range of ages -- \citet{morelli}, and also this paper. This diversity provokes a suggestion that
S0 galaxies may be formed through different evolutionary channels;
and many theoretical suggestions provide a variety of possible
physical mechanisms of S0 shaping (see Introduction). However,
properties of stellar populations, especially in the outer parts
of S0 galaxies less suffering from secular evolution (influencing
mainly the central parts of galaxies) can restrict strongly the 
possible scenaria of S0 galaxy formation. Namely, the old ages and
high magnesium-to-iron ratios of the large-scale stellar disks of
S0s exclude evidently their (trans-)formation from spirals at
intermediate redshifts, $z<0.5$. We have now in hand some hints on
the ages larger than 10 Gyr in many large-scale disks of S0 galaxies,
especially those populating dense environments: our sample studied in the present paper
reveals a majority of old disks, but it is very small and spread over
various environment types; and also the recent results by \citet{roediger} 
for the Virgo 53 S0s give the mean age of their (inner and) outer 
parts of $10.2 \pm 0.7$ Gyr. If the tendency for the outer disks of S0s
to be old is confirmed, all the scenaria which suggest spiral galaxy 
transformation into S0s in clusters and rich groups at intermediate 
redshifts would be disproved. What then remains? Only turbulent unstable
starforming thick gaseous disks at $z>1.5$ \citep{bournaud09} which
might leave thick quiescent stellar disks after a brief effective starforming
epoch, and perhaps also gas-rich major mergers, if they occur at $z>2$ 
in future cluster environments \citep{atlas3d_6}; 
and subsequent evolution with episodic rejuvenation according to our scenario 
proposed above. However, to make more certain conclusions, further investigations 
of the stellar population properties in the outer parts of lenticular galaxies 
are quite necessary.

\section*{Acknowledgments}

We thank V. L. Afanasiev, A. V. Moiseev, and S. S. Kaisin for fulfilling 
some of the observations which data are used in this work. The 6m telescope
is operated under the financial support of Science Ministry of Russia
(registration number 01-43). We are grateful to N. Ya. Sotnikova for proposing
a sample of edge-on lenticular galaxies for the observations. During our data 
analysis we used the Lyon-Meudon Extragalactic Database (LEDA) supplied by the
LEDA team at the CRAL-Observatoire de Lyon (France) and the NASA/IPAC
Extragalactic Database (NED) operated by the Jet Propulsion
Laboratory, California Institute of Technology under contract with
the National Aeronautics and Space Administration.
This research is partly based on SDSS data.
Funding for the Sloan Digital Sky Survey (SDSS) and SDSS-II has been
provided by the Alfred P. Sloan Foundation, the Participating Institutions,
the National Science Foundation, the U.S. Department of Energy, the National
Aeronautics and Space Administration, the Japanese Monbukagakusho,
and the Max Planck Society, and the Higher Education Funding Council
for England. The SDSS Web site is http://www.sdss.org/.
The study is supported by the grant of the Russian Foundation for Basic
Researches number 10-02-00062a.

\end{document}